\pdfoutput=1
\RequirePackage{ifpdf}
\ifpdf 
\documentclass[pdftex]{sigma}
\else
\documentclass{sigma}
\fi

\numberwithin{equation}{section}

\newtheorem{Theorem}{Theorem}[section]
\newtheorem{Corollary}[Theorem]{Corollary}
\newtheorem{Lemma}[Theorem]{Lemma}
 { \theoremstyle{definition}
\newtheorem{Remark}[Theorem]{Remark} }

\usepackage{mathtools}
\usepackage{tikz}
\usepackage{enumerate}
\usepackage{enumitem}

\newcommand{\R}{\mathbb{R}}
\newcommand{\Z}{\mathbb{Z}}

\newcommand{\C}{\mathbb{C}}
\newcommand{\T}{\mathbb{T}}
\newcommand{\M}{\mathbb{M}}

\newcommand{\be}{\begin{equation}}
\newcommand{\ee}{\end{equation}}
\newcommand{\beq}{\begin{equation}}
\newcommand{\eeq}{\end{equation}}
\newcommand{\bea}{\begin{eqnarray}}
\newcommand{\eea}{\end{eqnarray}}

\newcommand{\ol}{\overline}
\newcommand{\pa}{\partial}
\newcommand{\ti}{\tilde}

\newcommand{\si}{\sigma}
\newcommand{\la}{\lambda}

\newcommand{\id}{\mathbb{I}}
\newcommand{\I}{\mathrm{i}}
\newcommand{\E}{\mathrm{e}}

\newcommand{\sgn}{\mathop{\mathrm{sgn}}}
\newcommand{\clos}{\mathop{\mathrm{clos}}}
\newcommand{\re}{\mathop{\mathrm{Re}}}

\newcommand{\noprint}[1]{}

\begin{document}

\newcommand{\arXivNumber}{2012.12371}

\renewcommand{\PaperNumber}{045}

\FirstPageHeading

\ShortArticleName{How Discrete Spectrum and Resonances Influence the Asymptotics}

\ArticleName{How Discrete Spectrum and Resonances Influence\\ the Asymptotics of the Toda Shock Wave}

\Author{Iryna EGOROVA~$^{\rm a}$ and Johanna MICHOR~$^{\rm b}$}

\AuthorNameForHeading{I.~Egorova and J.~Michor}

\Address{$^{\rm a)}$~B.~Verkin Institute for Low Temperature Physics and Engineering,\\
\hphantom{$^{\rm a)}$}~47, Nauky Ave., 61103 Kharkiv, Ukraine}
\EmailD{\href{mailto:iraegorova@gmail.com}{iraegorova@gmail.com}}

\Address{$^{\rm b)}$~Faculty of Mathematics, University of Vienna,\\
\hphantom{$^{\rm b)}$}~Oskar-Morgenstern-Platz 1, 1090 Wien, Austria}
\EmailD{\href{mailto:Johanna.Michor@univie.ac.at}{Johanna.Michor@univie.ac.at}}
\URLaddressD{\url{http://www.mat.univie.ac.at/~jmichor/}}

\ArticleDates{Received January 21, 2021, in final form April 26, 2021; Published online May 01, 2021}

\Abstract{We rigorously derive the long-time asymptotics of the Toda shock wave in a~middle region where the solution is asymptotically finite gap. In particular, we describe the influence of the discrete spectrum in the spectral gap on the shift of the phase in the theta-function representation for this solution. We~also study the effect of possible resonances at the endpoints of the gap on this phase. This paper is a continuation of research started in~[arXiv:2001.05184].}


\Classification{37K40; 35Q53; 37K45; 35Q15}

\section{Introduction}

\begin{figure}[ht]
\centering
\includegraphics[width=11cm]{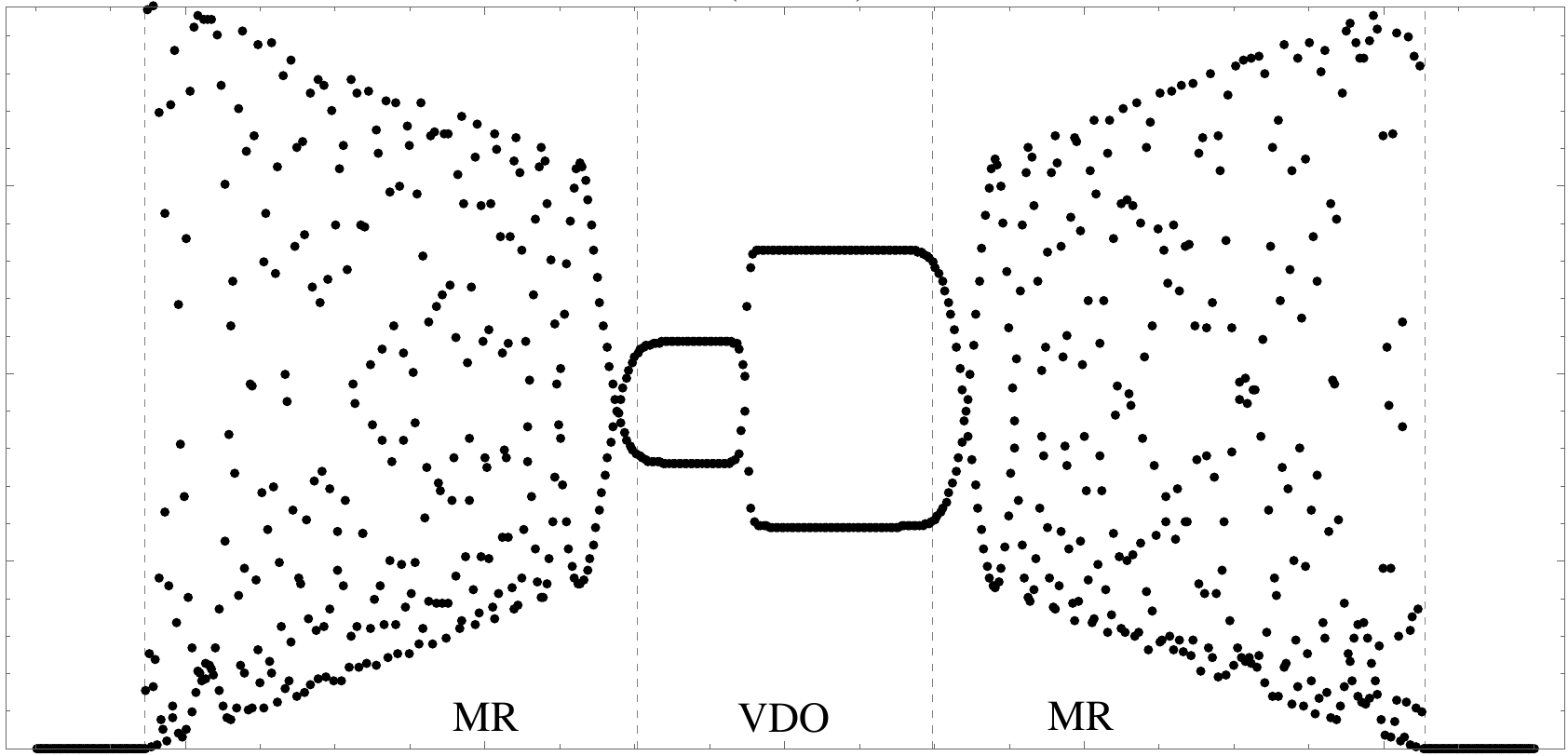}
\put(-327,0){\makebox(0,0)[lb]{\small $0.5$}}
\put(-327,36){\makebox(0,0)[lb]{\small $1.0$}}
\put(-327,73){\makebox(0,0)[lb]{\small $1.5$}}
\put(-327,110){\makebox(0,0)[lb]{\small $2.0$}}
\put(-175,151){\makebox(0,0)[lb]{\small $a(n,200)$}}
\put(-292,-9){\makebox(0,0)[lb]{\small $-400$}}
\put(-232,-9){\makebox(0,0)[lb]{\small $-200$}}
\put(-158,-9){\makebox(0,0)[lb]{\small $0$}}
\put(-104,-9){\makebox(0,0)[lb]{\small $200$}}
\put(-44,-9){\makebox(0,0)[lb]{\small $400$}}
\\[3ex]
\includegraphics[width=11cm]{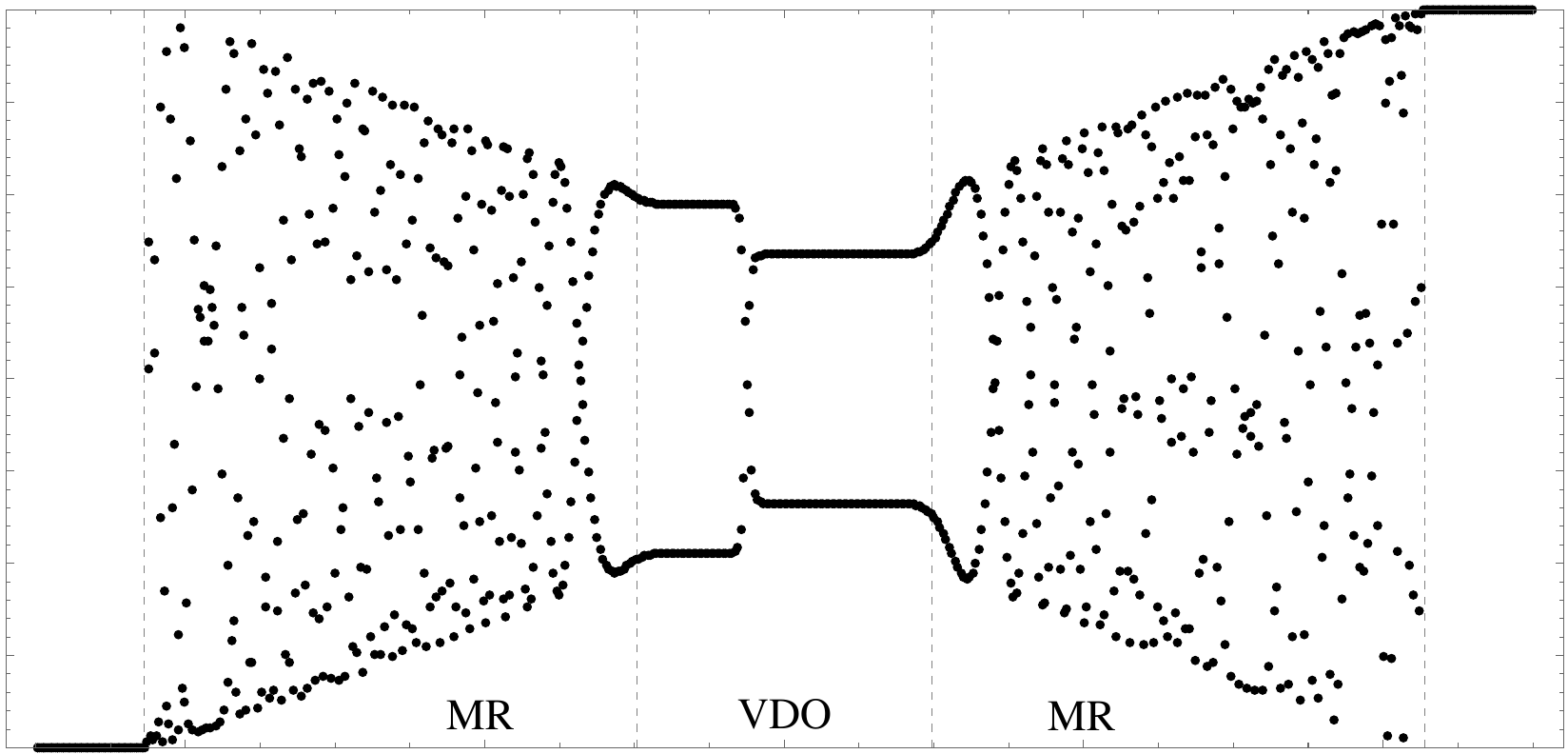}
\put(-335,-3){\makebox(0,0)[lb]{\small $-4.0$}}
\put(-335,15){\makebox(0,0)[lb]{\small $-3.5$}}
\put(-335,34){\makebox(0,0)[lb]{\small $-3.0$}}
\put(-335,53){\makebox(0,0)[lb]{\small $-2.5$}}
\put(-335,71){\makebox(0,0)[lb]{\small $-2.0$}}
\put(-335,89){\makebox(0,0)[lb]{\small $-1.5$}}
\put(-335,108){\makebox(0,0)[lb]{\small $-1.0$}}
\put(-335,126){\makebox(0,0)[lb]{\small $-0.5$}}
\put(-327,144){\makebox(0,0)[lb]{\small $0.0$}}
\put(-175,151){\makebox(0,0)[lb]{\small $b(n,200)$}}
\put(-292,-9){\makebox(0,0)[lb]{\small $-400$}}
\put(-232,-9){\makebox(0,0)[lb]{\small $-200$}}
\put(-158,-9){\makebox(0,0)[lb]{\small $0$}}
\put(-104,-9){\makebox(0,0)[lb]{\small $200$}}
\put(-44,-9){\makebox(0,0)[lb]{\small $400$}}
\caption{Numerically computed Toda shock wave with one eigenvalue.}\label{fig:num}
\end{figure}
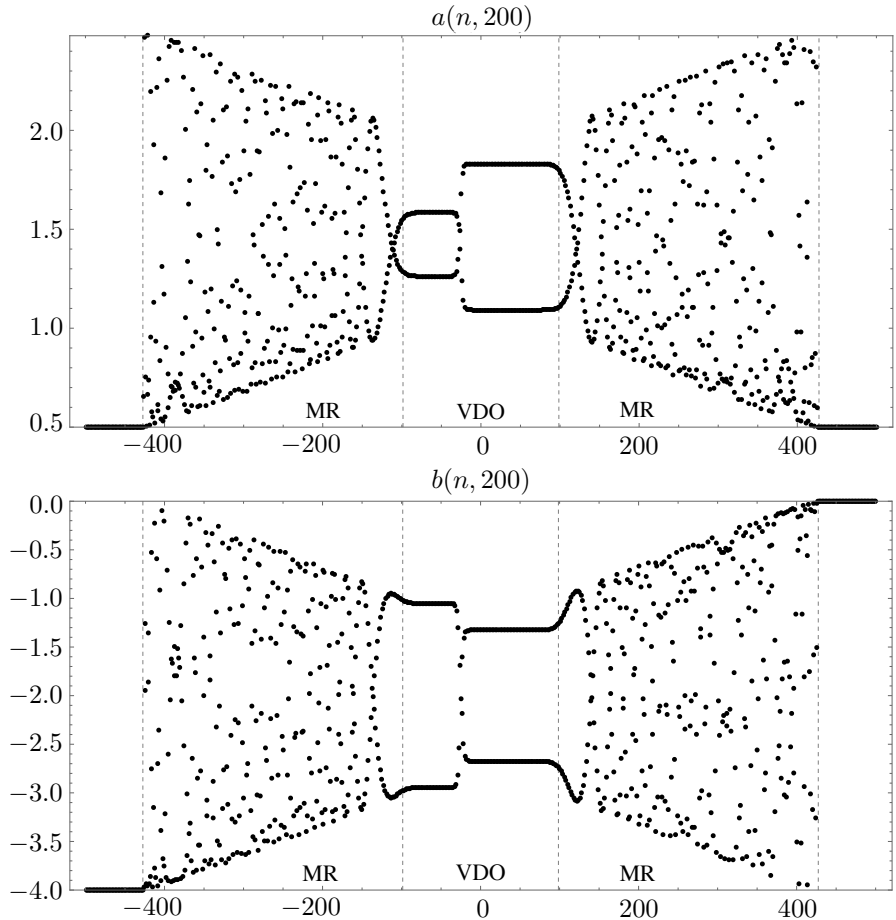

The Toda shock wave describes the motion of an infinite chain of particles with nonlinear nearest neighbor interactions
when the chain is excited with shock type initial conditions. We~are interested in the effect the eigenvalues in the spectral gap of the associated Lax operator have on the
asymptotic behavior of the shock wave.
The Toda shock wave is generated by the solution of the following initial value problem for the Toda lattice~\cite{tjac, toda}
\begin{gather}
\frac{\rm d}{{\rm d}t} \ti b(n,t) = 2\big(\ti a(n,t)^2 -\ti a(n-1,t)^2\big),\nonumber
\\[.5ex]
\frac{\rm d}{{\rm d}t} \ti a(n,t) = \ti a(n,t) \big(\ti b(n+1,t) -\ti b(n,t)\big),
\qquad (n,t) \in \Z \times \R_+,\label{tl}
\end{gather}
with a steplike initial profile $\big\{\ti a(n,0), \ti b(n,0)\big\}$ such that
\begin{gather} \label{ini1}
\ti a(n,0)\to a_\pm, \qquad
\ti b(n,0) \to b_\pm, \qquad
\text{as}\quad n \to \pm\infty,
\end{gather}
where $a_\pm >0$ and $ b_\pm\in\R$ satisfy the condition
\begin{gather} \label{main}
b_- + 2 a_- < b_+ -2a_+.
\end{gather}
This condition fixes the position of the background spectra relative to each other;
their mutual location produces essentially different types of asymptotic solutions~\cite{m15}.
The notion of the Toda shock wave~\cite{bk2, bk} was traditionally associated with symmetric initial data
\begin{gather}\label{symcoef}
\ti a(n-1,0)=\ti a(-n, 0),\qquad
\ti b(n,0)=-\ti b(-n,0),
\end{gather}
and the background constants $a_-=a_+=\frac{1}{2}$, $b_+=-b_->1$. The asymptotic of the solution of~\eqref{tl} for the particular case
\begin{gather}\label{steppure}
\ti a(n,0)=\frac{1}{2},\qquad
\ti b(n,0)=b \sgn n, \qquad
n\in\Z,\qquad \text{where}\quad \sgn 0=0,
\end{gather}
was studied in the pioneering work~\cite{vdo} by Venakides, Deift, and Oba in 1991. By use of the Lax--Levermore approach they established that in a middle region of the half plane $(n,t)\in\Z\times \R_+$, the asymptotic of the shock wave~\eqref{tl}, \eqref{steppure} is described by a 2-periodic solution of the Toda lattice. They also showed that the asymptotic undergoes a phase shift caused by the presence of a single eigenvalue at
$\la=0$. We~refer to this middle region of periodic asymptotics as VDO region,\footnote{Precise boundaries for the VDO region in our general case are given by~\eqref{gamma12}--\eqref{bound}.} compare Figure~\ref{fig:num}.

In this paper, we offer a derivation and rigorous justification of the asymptotic for~\eqref{tl}--\eqref{main} in the VDO region
using the vector Riemann--Hilbert problem (RHP) approach. We~allow more general initial data~\eqref{ini1} with arbitrary positive $a_\pm$ and $b_\pm$ satisfying~\eqref{main}. In particular,
the novel features are:
\begin{itemize}\itemsep=0pt
\item an arbitrary discrete spectrum,

\item possible resonances at the edges of the continuous spectrum,

\item no symmetry assumption~\eqref{symcoef},

\item a partial revision of results in~\cite{vdo} including estimates on the error terms,

\item a finite gap (two band) asymptotic due to spectra of different length.
\end{itemize}

\begin{figure}[ht]
\centering
\begin{tikzpicture}

\fill[gray!5] (0,0) -- (-2.7,4.23) -- (-2.9,4.1) --(0,0);
\draw[thick, dotted] (0,0) -- (-2.7,4.23);
\fill[gray!5] (0,0) -- (-4.4,3.3) -- (-4.2,3.4) --(0,0);
\draw[thick, dotted] (0,0) -- (-4.4,3.3);
\fill[gray!5] (0,0) -- (1.15,5) -- (0.9,5) --(0,0);
\draw[thick, dotted] (0,0) -- (1.15,5);
\fill[gray!5] (0,0) -- (-0.2,5) -- (0.1,5) --(0,0);
\draw[thick, dotted] (0,0) -- (-0.2,5);

\draw[->, thick] (-5,0) -- (5,0) node[right] {$n$};
\draw[->, thick] (0,0) -- (0,4.5) node[above] {$t$} ;

\fill[gray!5] (0,0) -- (3.9,4.3) -- (1.6,4.8) -- (0,0);
\fill[gray!5] (0,0) -- (-0.6,4.8) -- (-2.3,4.3) -- (0,0);

\node at (2,3.5) {$ $};
\node at (-1.1,3.5) {$\frac n t \in \mathcal I_\varepsilon^j$};
\node at (2,3.5) {$\frac n t \in \mathcal I_\varepsilon^1$};

\draw (0,0) -- (4,4) node[right] {$\tfrac{n}{t}=\xi_0$};
\draw[thick, dotted] (0,0) -- (3.9,4.3) node[above right] {$\tfrac{n}{t}=\xi_0-\varepsilon$};
\draw[thick, dotted] (0,0) -- (1.6,4.8) node[right] {$\tfrac{n}{t}=\xi_1+\varepsilon$};
\draw (0,0) -- (1.4,5) node[above] {$ $};
\draw (0,0) -- (-0.4,5) node[above] {$\tfrac{n}{t}=\xi_{j-1}$};
\draw[thick, dotted] (0,0) -- (-0.6,4.8);
\draw[thick, dotted] (0,0) -- (-2.3,4.3);
\draw (0,0) -- (-2.5,4.3) node[above] {$\tfrac{n}{t}=\xi_j$};
\draw (0,0) -- (-4.5,3.2) node[above] {$\tfrac{n}{t}=\xi_{\aleph +1}$};

\end{tikzpicture}
\caption{The VDO region.} \label{fig:2}
\end{figure}
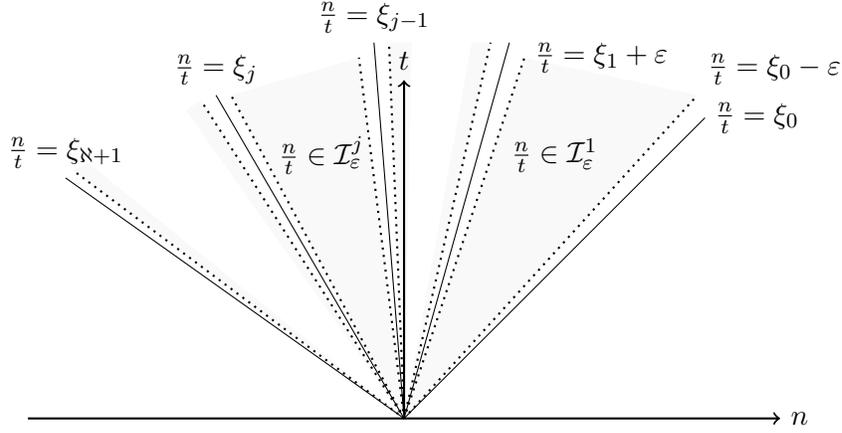

The vector RHP approach in the context of the Toda problem was proposed in~\cite{dkkz} and further developed in~\cite{bt, emt14, km0, KTa, KTb}. We~use standard conjugations/deformations such as the $g$-function technique~\cite{dvz} which proved its efficiency in steplike cases. A~suitable $g$-function for the VDO region replaces the standard phase function and makes it possible to apply the lense mechanism. It~also provides a characterization of the boundaries of the sectors (see Figure~\ref{fig:2}) where the asymptotics are given by a finite gap solution of~\eqref{tl} with unaltered phase. We~des\-cribe the $g$-function for the VDO region as an Abel integral
on the Riemann surface associated with the continuous two band spectrum of the underlying Jacobi operator of~\eqref{tl} in Section~\ref{sec3}.

Before we state our main theorem, let us first note that without loss of generality it is sufficient to study the case of background spectra $[b-2a, b +2a]\cup [-1,1]$. Indeed, assume that the vector-function
$\big(\ti a(t),\ti b(t)\big)= \big\{\ti a(n,t), \ti b(n,t)\big\}_{n\in\Z}$ is the solution of~\eqref{tl}--\eqref{main}. Then the function $(a(t), b(t))$ given by
\begin{gather*}
a(n,t)=\frac{1}{2 a_+}\ti a\bigg(n, \frac{t}{2 a_+}\bigg), \qquad
b(n,t)=\frac{1}{2 a_+}\ti b\bigg(n,\frac{t}{2a_+}\bigg) - b_+,\qquad
n\in \Z,
\end{gather*}
satisfies the initial value problem
\begin{gather}
\frac{\rm d}{{\rm d}t} b(n,t) = 2\big( a(n,t)^2 - a(n-1,t)^2\big),\nonumber
\\
\frac{\rm d}{{\rm d}t} a(n,t) = a(n,t) ( b(n+1,t) - b(n,t)),
\qquad (n,t) \in \Z \times \R_+,\label{tl1}
\\
a(n,0)\to \frac 1 2, \quad b(n,0)\to 0,\quad n\to +\infty; \qquad
a(n,0)\to a,\quad b(n,0)\to b,\quad n\to -\infty,\nonumber
\end{gather}
with{\samepage
\begin{gather}\label{main3}
b+2a<-1,
\end{gather}
where we denoted
$b:=b_- - b_+$, $a:=\frac{a_-}{2 a_+}$.}

Hence it suffices to study the shock wave~\eqref{tl1}--\eqref{main3}. We~assume
that the initial data tend to the background constants exponentially fast with some small rate $\rho>0$,
\begin{gather} \label{decay}
\sum_{n=1}^\infty \E^{\rho n}\bigg(\bigg|a(n,0)-\frac 1 2 \bigg| + |b(n,0)| +|a(-n,0)-a| + |b(-n,0)-b|\bigg)<\infty.
\end{gather}

Figure~\ref{fig:num} demonstrates the behavior of the Toda shock wave corresponding to the initial data $a(n,0)=\frac 12$, $n\in\Z$; $b(n,0)=-4$, $n<0$; $b(0,0)=-1.7$, $b(n,0)=0$, $n>0$, at a large but fixed time $t=200$. Such initial data have one eigenvalue in the gap and the background spectra are of~equal length. Hence the asymptotic of the shock wave in the VDO region is periodic with period~$2$ and exhibits one phase shift. In the left and right modulation regions (MR) the
asymptotic is a modulated single-phase quasi-periodic Toda solution as discussed in~\cite{empt19}.\vspace{0.5ex}

The initial data~\eqref{decay} can have a finite discrete spectrum. We~enumerate the eigenvalues $\la_j$ in~the gap $(b+2a, -1)$ increasingly starting from the leftmost; let $\aleph$ be the number of eigenvalues
in the gap. Given an arbitrary\footnote{The maximal value of $\varepsilon$ which is admissible for our purpose is
specified in Section~\ref{Sec:4}.} small $\varepsilon>0$, the VDO region consists of $\aleph +1$ disjoint regions
\begin{gather*}
\bigg \{(n,t)\colon \frac{n}{t}\in \mathcal I_\varepsilon^j=[\xi_{j}+\varepsilon, \xi_{j-1}-\varepsilon]\bigg\}
\end{gather*}
as depicted in Figure~\ref{fig:2},
where $\xi_j$ are the points $\xi$ at which the level line $\re g(\la, \xi)=0$ of the $g$-function (cf.~Section~\ref{sec3}) crosses $\R$ at $\la_j$. We~denote by $\xi_0$ and $\xi_{\aleph+1}$ the points where $\re g(b+2a,\xi_0)=0$ and $\re g(-1,\xi_{\aleph+1})=0$, and the rays $\tfrac{n}{t}=\xi_{\aleph +1}$ and $\tfrac{n}{t}=\xi_0$ determine the outer boundaries of the VDO region. With each interval $\mathcal I_\varepsilon^j$ we associate a shift phase $\Delta_j$ (not depending on $\varepsilon$) expressed in terms of the initial scattering data for the solution of~\eqref{tl1} and~\eqref{decay} (see~\eqref{Deltaj} below). For~each $\Delta_j$ one finds via Jacobi's inversion problem~\eqref{jacinv} the initial Dirichlet divisor
 and the unique finite gap solution $\big\{\hat a(n,t,\Delta_j), \hat b(n,t,\Delta_j)\big\}$ from the isospectral set associated with the two-band spectrum $\mathfrak S:=[b-2a, b+2a]\cup[-1,1]$.
Our main result is

\begin{Theorem}\label{theor:main}
Let $\{a(n,t), b(n,t)\}$ be the solution of the initial value problem~\eqref{tl1}--\eqref{main3},~\eqref{decay} and let $n\to\infty$, $t\to\infty$ with $\frac{n}{t}\in \mathcal I_\varepsilon^j$, where $\varepsilon>0$ is an arbitrary, sufficiently small number. Let~$\big\{\hat a(n,t,\Delta_j), \hat b(n,t,\Delta_j)\big\}$ be the finite gap solution associated with the spectrum $\mathfrak S$ and with the phase $\Delta_j$ given by~\eqref{def ab}, \eqref{Deltaj}, \eqref{zavis}, \eqref{defmatp}, \eqref{defQuu}, \eqref{defchi}. Then there exists $C(\varepsilon)>0$ such that
\begin{gather}\label{estim}
b(n,t)=\hat b(n,t,\Delta_j) + O\big(\E^{-C(\varepsilon)t}\big), \qquad
a(n,t)=\hat a(n,t,\Delta_j) + O\big(\E^{-C(\varepsilon)t}\big).
\end{gather}
 The shift of the phases at the point $\la_j=\frac{z_j+z_j^{-1}}{2}\in (b+2a, -1)$ of the discrete spectrum is given~by
 \begin{gather*}
\Delta_j-\Delta_{j+1}=2 \frac{\int_{q_1}^q \log \left|\frac{z_j s - 1}{s-z_j}\right| \big|\ti{\mathcal P}^{-1}(s)\big|{\rm d}s}{\int_{-1}^{q_1} \ti{\mathcal P}^{-1}(s)\,{\rm d}s}, \qquad\!
\ti{\mathcal P}(s)=\sqrt{(s-q)(s-q_1)\big(s-q_1^{-1}\big)\big(s-q^{-1}\big)},
\end{gather*}
where $q$ and $q_1$ are defined by~\eqref{simpl4}.
\end{Theorem}

\begin{Remark}
$(i)$ For initial data~\eqref{decay} the scattering data consist of the modulo of the right transmission coefficient $|T(\la)|$ given on $[b-2a, b+2a]$, the right reflection coefficient $R(\la)$ on~$[-1,1]$ and the discrete spectrum on
$\R\setminus( [b-2a, b+2a]\cup [-1,1])$. As expected, we see that $R(\la)$ and the discrete spectrum to the right of $\la_j$ do not influence the asymptotic in the sector $\frac{n}{t}\in\mathcal I_\varepsilon^j$.

$(ii)$ The error terms in~\eqref{estim} are of order $O\big(\E^{-C(\varepsilon)t}\big)$ and thus significantly better than the
estimate $O\big(t^{-1}\big)$ one would expect by analogy with the error estimates in the modulation re\-gions~\cite{empt19}.
The error terms in~\eqref{estim} were obtained by a careful analysis of the relations between the analytic continuation of the scattering functions. These relations allowed us to prove that there are no parametrix points~\cite{dkmvz} in the RHP for the VDO region.

$(iii)$ We use vector RHP statements instead of~matrix statements (as do \cite{aelt16,egkt13,GGM,grunert_teschl,piork19} in the
case of the KdV equation with steplike initial data), because the matrix statements for the shock wave are ill-posed
for certain arbitrary large values of~$n$ and~$t$ in~the class of~invertible matrices with $L^2$-integrable singularities on the jump contour, for both the initial and model RHPs.
This fact for Toda shock can be established similarly as for the KdV case~\cite{ept19}. One would have to admit
then additional poles for solutions outside the discrete spectrum in~the matrix statements. This makes proving
uniqueness of~the solutions far more difficult. The statements of~the RHPs in~vector form
together with additional symmetries to be posed on contours, jump matrices and on the solutions itself imply
uniqueness almost straightforward. However, for the final small-norm arguments we need to construct an
invertible matrix model RHP solution. It~has poles and it might not be unique, but the corresponding error vector function has no poles. Such a solution is given in~Lemma~\ref{aboutmatrix}.

$(iv)$ Unlike to KdV, for the Toda equation with non-overlapping background spectra, the statements of~the RHPs associated with left and right initial data look identical. The proper choice of~the initial statement for the RHP can essentially simplify the further analysis in a~given region of~space-time variables $\Z\times \R_+$ (cf.~\cite{emt14}). For~the VDO region both choices are appropriate.
\end{Remark}

\section{Notations and statement of the initial holomorphic RHP} 
To maintain generality of the presentation while keeping notations short, we formulate all preliminary facts on the inverse scattering transform for the steplike initial profile in terms of the spectral variables $z_\pm$ associated with the initial data~\eqref{ini1}, \eqref{main}. First of all, let us list some well known properties of the scattering data for the steplike Jacobi operator $\ti H(t)$ involved in the Lax representation $\frac{\rm d}{{\rm d}t}\ti H(t)=\big[\ti H(t), \ti A(t)\big]$ for the initial value problem~\eqref{tl}, \eqref{ini1}.~This~prob\-lem has a unique solution (cf.~\cite{tjac}). Assume that the coefficients of the initial Jacobi operator $\ti H(0)$ tend to the limiting (or background) constants $a_\pm$, $b_\pm$ with a first summable moment of~perturbation, that is, $n(\ti a(n,0) - a_\pm)\in\ell^1(\Z_\pm)$ and $n(\ti b(n,0) - b_\pm)\in\ell^1(\Z_\pm)$. Then the unique solution $\big\{\ti a(t), \ti b(t)\big\}$ of~\eqref{tl} satisfies
\begin{gather}\label{firstmom}
n\big(\ti a(n,t) - a_\pm\big)\in\ell^1(\Z_\pm), \qquad
n\big(\ti b(n,t) - b_\pm\big)\in\ell^1(\Z_\pm).
\end{gather}
 With this condition fulfilled, introduce some notations and notions.
 \begin{itemize}
\item {\it The background} Jacobi operators
\begin{gather*}
H_\pm y(n):=a_\pm y(n-1)+ b_\pm y(n) +a_\pm y(n+1),\qquad n\in\Z,
\end{gather*}
have spectra $\si_\pm =[ b_\pm-2a_\pm, b_\pm+ 2 a_\pm]$ which do not overlap, and by~\eqref{main} satisfy \mbox{$\sup \si_-<\inf\si_+$}.
\item {\it The Joukovski maps} $z_\pm=z_\pm(\la)$ of the spectral parameter $\la$ are given by
\begin{gather*}
\la= b_\pm +a_\pm \big(z_\pm + z^{-1}_\pm\big),\qquad
z_\pm\colon\ \clos(\C\setminus \si_\pm)\mapsto |z_\pm|\leq 1.
\end{gather*}
The map $z_+\mapsto z_-$ is one-to-one between the domains $\mathcal D_+$ and $\mathcal D_-$, where \begin{gather*}
\mathcal D_\pm=\{z_\pm\colon |z_\pm|<1, z_\pm \notin z_\pm(\si_\mp)\}.
\end{gather*}
 The functions $\big\{z_\pm^{\pm n}\big\}_{n\in\Z}$ are called the {\it free exponents}. They solve the background spectral equations $H_\pm y(n)=\la y(n)$.
\item The operator $\ti H(t)$ has an absolutely continuous spectrum on the set $\si_+\cup \si_-$ and
a finite discrete spectrum $\si_{\rm d}$, which we divide into three parts,
\begin{gather*}
\si_{\rm d}^{\rm left}=\{\la_j\in\si_{\rm d}\colon \la_j<b_- -2a_-\},\qquad \si_{\rm d}^{\rm right}=\{\la_j\in\si_{\rm d}\colon \la_j>b_++2a_+\},
\\
\si_{\rm d}^{\rm gap}=\{\la_j\in\si_{\rm d}\colon b_-+2a_-<\la_j<b_+-2a_+\}.
\end{gather*}
 The points $z_j^\pm=z_\pm(\la_j)\in \mathcal D_\pm\cap(-1,1)$, $\la_j\in\si_{\rm d}$, are also called points of the discrete spectrum.
\item {\it The Jost solutions} of the spectral equation
\begin{gather*}
\ti a(n-1,t) \psi^\pm(\la, n-1, t) +\big( \ti b(n,t)-\la\big)\psi^\pm(\la, n, t) +\ti a(n,t)\psi^\pm(\la, n+1, t)=0
\end{gather*}
are normalised as
\begin{gather*}
\lim_{n\to\pm\infty}(z_\pm)^{\mp n}\psi^\pm(\la, n, t)=1,\qquad
\la\in
\ \clos(\C\setminus \si_\pm).
\end{gather*} We can consider them as functions of $z_\pm$ in the closures of
$\mathcal D_\pm$. Their {\it Wronskian}
\begin{gather}\label{wrons}
\ti W(\la):=a(n-1,0)\big(\psi_-(\la, n-1, 0)\psi_+(\la, n,0) -\psi_+(\la, n-1, 0)\psi_-(\la,n,0)\big)
\end{gather}
is an important spectral characteristic of the steplike scattering problem. It~can be treated as an analytic function of $z_\pm$ in $\clos \mathcal D_\pm$.
The points $z_j^\pm$ are its simple zeros, and $\psi_\pm(\la_j,n,t)$ are the (dependent) eigenfunctions.
\item {\it The normalising constants} are introduced by
\begin{gather*}
\bigg(\sum_{n\in\Z}\psi_\pm^2(\la_j,n,t)\bigg)^{-2}=\gamma_j^\pm(t)
=\gamma_j^\pm(0)\E^{z_j^\pm -(z_j^\pm)^{-1}}.
\end{gather*}
\item
{\it The scattering relations}
\begin{gather}\label{scatrel}
T_\pm(\la,t)\psi_\mp(\la,n,t)=R_\pm(\la,t)\psi_\pm(\la,n,t) +\overline{\psi_\pm(\la,n,t)}
\end{gather}
hold on the sets $|z_\pm|=1$.
 \item {\it The time evolution} of the scattering data is given by
\begin{alignat}{3}
& R_\pm(\la,t)=R_\pm(\la,0)\E^{\pm z_\pm\mp z_\pm^{-1}}\qquad&& \text{for}\quad |z_\pm|=1,& \nonumber
\\
&|T_\pm(\la,t)|^2=|T_\pm(\la,0)|^2\E^{\pm z_\mp\mp z_\mp^{-1}}\qquad&& \text{for}\quad |z_\mp|=1.&\label{evolu}
\end{alignat}
\end{itemize}
Relations~\eqref{scatrel} and~\eqref{evolu} hold under condition~\eqref{firstmom} which guarantees existence and good analytical properties of the Jost solutions. The function $|T_\pm(\la,0)|^2$ cannot be continued analytically outside the domain $|z_\mp|=1$. However, if the initial data tend to the limiting constants exponentially fast with a rate $\rho>0$ (cf.~\eqref{decay}), then the right hand side of the $\pm$-scattering relation continues analytically in the domain $1-\rho<|z_\pm|\leq 1$, and the respective equality~\eqref{scatrel} is preserved. In particular, the reflection coefficient $R_\pm(\la,0)$ continues in the domain $1-\rho<|z_\pm|\leq 1$, and the function $\chi(\la)$ defined in~\eqref{defchi} below
can be continued analytically in both domains.

The vector RHP connected with the scattering problem for $\ti H(t)$ can be stated in two ways, based either
on the right or left scattering data. The correct choice of the scattering data which significantly simplifies the further analysis depends on the region of the $(n,t)$ half-plane for which the asymptotic of~\eqref{tl}--\eqref{main} should be derived.
In our situation, the VDO region on the $\Z\times \R_+$ half plane could be analysed via left or right RHP and both cases are equivalent in structure and complexity of steps.
To state a proper vector RHP we proceed as follows.

Let $\mathbb M$ be the two-sheeted Riemann surface associated with the function
\begin{gather*}
w(\la)=\big(\big((\la - b_-)^2 - 4(a_-)^2\big)\big((\la - b_+)^2 - 4(a_+)^2\big)\big)^{1/2},
\end{gather*}
with glued cuts along $\si_+$ and $\si_-$. Denote a point on $\mathbb M$ by $p=(\la,\pm)$.
On the upper sheet of $\mathbb M$ introduce two $1\times 2$ vector-functions $M^{\pm}(p,n,t):=M^{\pm}(p)$ (here variables $n$ and $t$ are treated as parameters) by
\begin{gather*}
M^{\pm}(p)=\big(T_\pm(\la,t)\psi_\mp(\la,n,t)(z_\pm(\la))^n, \psi_\pm(\la,n,t))(z_\pm(\la))^{-n}\big),\qquad p=(\la,+).
\end{gather*}
The first component of each function is a meromorphic function of $p$ on the upper sheet on $\mathbb M$ with simple poles at points of $\si_{\rm d}$ and known residues. At infinity $M^\pm(p)$ have finite values and the product of components is equal to 1~\cite{emt14}.
 Let us extend each function $M^\pm$ to the lower sheet~by
\begin{gather}\label{symini}
M^{\pm}(p^*)=M^{\pm}(p)\si_1,
\end{gather}
where
$\si_1=\left( \begin{smallmatrix}0&1\\1&0\end{smallmatrix} \right)$
is the first Pauli matrix and $p^*=(\la,-)$ is the involution point for $p=(\la,+)$.
With this extension, both functions have jumps along the boundaries of the sheets on $ \mathbb M$, and these jumps can be easily evaluated. The jump problems together with normalisation conditions $M^\pm_1(\infty_+)M_1^\pm(\infty_-)=1$, residue conditions at points of $\si_{\rm d}$ and $\si_{\rm d}^*$, and symmetry condition~\eqref{symini}, form the content of the
left and right RHPs associated with~\eqref{tl}--\eqref{main}.

In this paper, we use the traditional RHP statement based on the right scattering data in~terms of the variable $z_+$.
As discussed in the introduction, we restrict ourselves to the case
\begin{gather*}
a_+=\frac{1}{2}, \qquad
b_+=0,\qquad
a_-=a>0,\qquad
b_-=b,\qquad
b+2a<-1,
\end{gather*}
and denote by $\mathfrak{S}=[b-2a, b+2a]\cup [-1, 1]$ the continuous spectrum of the Jacobi operator involved in~\eqref{tl1}.
To ease notations, we omit from here on the subscript ``$+$" in the notations and set
\begin{gather}
 z(\la):=z_+(\la),\qquad \la=\frac{z+z^{-1}}{2},\qquad |z|\leq 1, \nonumber
\\
\label{simpl4}
 q=z(b-2a),\qquad q_1=z(b+2a),\qquad z_j:=z_+(\la_j),\qquad \gamma_j:=\gamma_j^+(0).
\end{gather}

\begin{Remark} We use the formal notation $z_j\in\si_{\rm d}$, $\si_{\rm d}^{\rm gap}$ if $\la_j=\frac{z_j +z_j^{-1}}{2}\in\si_{\rm d}$,
$\si_{\rm d}^{\rm gap}$, respectively. Let us enumerate the points $z_j$ starting from $\si_{\rm d}^{\rm gap}$, that is,
\begin{gather*}
-1<z_\aleph<\dots<z_1<q_1,\qquad \aleph=\text{Card}\, \si_{\rm d}^{\rm gap}.
\end{gather*}
All remaining points of the discrete spectrum will lie outside of $\si_{\rm d}^{\rm gap}$.
\end{Remark}
Further notations are
\begin{gather}
R(z):=R_+(\la,0) \qquad \text{for}\quad |z|=1, \nonumber
\\
\psi(z,n,t)=\psi_+(\la,n,t) \qquad \text{for}\quad |z|\leq 1,\nonumber
\\
T(z,t):=T_+(\la,t),\qquad \psi_{{\rm left}}(z,n,t)=\psi_-(\la(z),n,t) \qquad \text{for}\quad z\in \mathcal D,\label{simpl3}
\end{gather}
 where
 \begin{gather}\label{defmathcald}
 \mathcal D=\{z\colon |z|<1,\, z\notin[q_1,q]\}.
 \end{gather}
 The domain $\mathcal D$ is in one-to-one correspondence with the upper sheet of $\mathbb M$ (we treat sheets as open sets) with
 \begin{gather*}
 z =\la-\sqrt{\la^2 -1} \leftrightarrow p=(\la, +), \qquad z\in \mathcal D.
 \end{gather*}
The domain
 \begin{gather}\label{defdstar}
 \mathcal D^*=\big\{z\colon z^{-1}\in \mathcal D\big\}
 \end{gather}
 corresponds to the lower sheet by
 \begin{gather*}
 z^{-1}=\la +\sqrt{\la^2 -1} \leftrightarrow p^*=(\la,-).
 \end{gather*}
 Therefore, the meromorphic RHP for $M^+(p)$ on $\mathbb M$ can be reformulated as an equivalent meromorphic RHP for $m(z)=M^+(p(z))$ on the $z$-plane, with jumps along the unit circle $\mathbb T=\{z\colon |z|=1\}$ and intervals $[q_1, q]$ and $\big[q^{-1}, q_1^{-1}\big]$. In this paper, we propose a slightly different (holomorphic) statement of the initial RHP, which is equivalent to the RHP for $M^+(p)$ on $\mathbb M$, and therefore has a unique solution (cf.~\cite{emt14}). This statement is specific for the domain VDO, where we derive the asymptotics, and allows us to skip several of the standard transformations, such as the reformulation of the meromorphic problem as a holomorphic problem and one of~two steps corresponding to opening of lenses.

 Let us choose a large natural number $N\gg 1$ and set
 \begin{gather}\label{defeps}
 \delta\leq N^{-1}\min_{z_i\neq z_j\in\si_{\rm d}}\big\{|z_i-q|, |z_i - z_j|, |z_i - q_1|, ||z_i|-1|\big\}.
 \end{gather}
We can always assume that $\delta<\rho$, where $\rho$ is the decay rate from~\eqref{decay}.
 Then the right Jost solutions $\psi(z,n,t)$ and $\psi(z^{-1}, n,t)$ are holomorphic functions in an $\delta$-vicinity of the unit circle~$\mathbb T$, and the standard scattering relation
 \begin{gather*}
 T(z,t)\psi_{{\rm left}}(z,n,t)=\psi\big(z^{-1},n,t\big) + R(z,t)\psi(z,n,t)
 \end{gather*}
 is continued analytically in the open ring
 \begin{gather*}
 \ti \Omega_\delta=\{z\colon 1-\delta<|z|<1\}.
 \end{gather*}
 With our choice of $\delta$, there are no points of the discrete spectrum in $\ti \Omega_\delta$, moreover,
 \begin{gather*}
 \inf_{z_j\in\si_{\rm d}}\mathrm{dist}(z_j, \mathcal C_\delta)>(N-1)\delta,
 \end{gather*}
 where we denoted
 \begin{gather}\label{Ceps}\mathcal C_\delta=\{z\colon |z|=1-\delta\}.\end{gather}
In particular, the continuation of the initial reflection coefficient $R(z)$ is an analytic function in~$\ti \Omega_\delta$.
Set
\begin{gather}
\mathbb D_{\delta,j}=\{z\colon |z-z_j|<\delta\}, \qquad
\mathbb T_{\delta,j}=\{z\colon |z-z_j|=\delta\},\qquad
\mathbb D_{\delta,j}^*=\big\{z\colon z^{-1}\in\mathbb D_{\delta,j}\big\},\nonumber
\\
\mathcal D_\delta=\mathcal D\setminus\Big(\ol{\ti\Omega_\delta\cup\bigcup_{\si_{\rm d}}\mathbb D_{\delta,j}}\Big),\qquad \mathcal D_\delta^*=\big\{z\colon z^{-1}\in\mathcal D_\delta\big\}.\label{circles}
\end{gather}
In $\ol{\mathcal D\setminus\Sigma_\delta},$ where
\begin{gather}\label{sig}
\Sigma_\delta=I\cup \mathcal C_\delta\cup\bigcup_{\si_{\rm d}}\mathbb T_{\delta,j},
\end{gather}
with $I:=[q_1,q]$, introduce the vector-function $m(z)=(m_1(z,n,t), m_2(z,n,t))$ by\footnote{This is a function of $z$, and $n$ and $t$ are treated as large parameters.}
\begin{gather}
\label{m_ini}
m(z)=\begin{cases}
\big(T(z,t)\psi_{{\rm left}}(z,n,t)z^n, \psi(z,n,t)z^{-n}\big),& z\in\mathcal D_\delta,
\\
\big(\psi(z^{-1},n,t) z^n, \psi(z,n,t)z^{-n}\big),& z\in \ti\Omega_\delta,
\\
\big(T(z,t)\psi_{{\rm left}}(z,n,t)z^n, \psi(z,n,t)z^{-n}\big)A_j(z),& z\in \mathbb D_{\delta, j}.
\end{cases}
\end{gather}
Here
\begin{gather}
\label{A_j}A_j(z)=
\begin{pmatrix}
1& 0
\\
(z-z_j)^{-1}\gamma_j z_j^{2n+1}\E^{t(z_j -z_j^{-1})}&1
\end{pmatrix}\!.
\end{gather}

\begin{Lemma}[\cite{emt14}]
\label{lem:as} We have
\begin{gather}\label{asympm}
m_1(0,n,t)=\prod_{k=n}^\infty 2 a(k,t), \qquad
\lim_{z\to 0}\frac{1}{2z}(m_1(z,n,t)m_2(z,n,t)-1)=b(n,t).
\end{gather}
\end{Lemma}
Extend $m(z)$ to $\mathcal D^*\setminus \Sigma^*_\delta$ with $\Sigma^*_\delta=\big\{z\colon z^{-1}\in\Sigma_\delta\big\}$ by
\begin{gather}\label{symcon}
m(z^{-1})=m(z)\sigma_1.
\end{gather}
Formula~\eqref{symcon} implies that the vector function~\eqref{m_ini}, considered as a piecewise-analytic function in $\mathbb C$, has jumps along the circle $\mathcal C_\delta$, along the interval $I$ and the small circles $\mathbb T_{\delta,j}$, as well as along their images $\mathcal C_\delta^*$, $I^*$ and $\mathbb T_{\delta,j}^*$ under the map $z\to z^{-1}$. However, $m(z)$ does not have a~jump along the unit circle $|z|=1$, i.e., it is holomorphic in the ring $1-\delta<|z|<(1-\delta)^{-1}$. The fact that $m(z)$ does not have any singularities at $z_j\in\si_{\rm d}$ is established in~\cite{dkkz} and~\cite{KTb}.

The {\it symmetry condition}~\eqref{symcon} plays a crucial role in establishing uniqueness of the solution for RHPs, and we cannot violate it. For~this reason, the initial RHP and all its further transformations (deformations and conjugations) should satisfy the following {\it symmetry constraints} $(i)$ and $(ii)$. Let $\Sigma$ be the jump contour of a generic RHP.
\begin{enumerate}\itemsep=0pt
 {\it \item[$(i)$] The jump contour $\Sigma$ should be symmetric with respect to the map $z\mapsto z^{-1}$, i.e., with every point $z$ it also contains $z^{-1}$.}
{\it \item[$(ii)$] Symmetric parts of $\Sigma$ are oriented in such a way that the jump matrix $\ti v(z)$ of the problem $\ti m_+(z)=\ti m_-(z)\ti v(z)$ and the solution itself satisfy the symmetries }
 \begin{gather*}
 \ti v(z)=\si_1 \ti v\big(z^{-1}\big)\si_1, \qquad z\in\Sigma;\qquad
 \ti m(z)= \ti m\big(z^{-1}\big)\si_1,\qquad z\in\C\setminus\Sigma.
 \end{gather*}
\end{enumerate}
 {\it Constraint $(ii)$} implies that the orientation of symmetric parts of $\Sigma$ is as follows: if a point~$z$ moves along a part of the contour $\mathcal K\subset\Sigma\cap\{z\colon |z|<1\}$ in the positive direction, then the point~$z^{-1}$ moves simultaneously in the positive direction of the symmetric part $\mathcal K^*$, where~$\mathcal K\cap\mathcal K^*=\varnothing$.
Except for the lense mechanisms where triangle matrices are used, all conjugations of the solution vector consist of multiplication by diagonal matrices of the form $[d(z)]^{-\sigma_3}$, that is, in transformations
$\ti m(z)\mapsto \ti m(z)[d(z)]^{-\sigma_3}$, where $d\colon \C\setminus \Sigma\to\C$ is a sectionally analytic function and $\sigma_3=(\begin{smallmatrix} 1&0\\0&-1\end{smallmatrix})$.
On all such conjugations we pose the {\it symmetry constraint} $(iii)$:
\begin{enumerate}\itemsep=0pt
{\it \item[$(iii)$] The contour $\Sigma$ of a non-analyticity for $d(z)$ should be symmetric with respect to $z\mapsto z^{-1}$. Moreover, the function $d(z)$ should satisfy either the property
\begin{gather*}
d\big(z^{-1}\big)=d(z)^{-1},\qquad z\in\C\setminus\Sigma,
\end{gather*}
or the property}
\begin{gather*}
d\big(z^{-1}\big)=d(z)\neq 0,\qquad z\in\C\setminus\Sigma, \qquad d(0)=1.
\end{gather*}
\end{enumerate}
Recall that $m(z)$ in~\eqref{m_ini} has bounded positive limits of both components at $0$ and $\infty$, moreover, by Lemma \ref{lem:as},
\begin{gather}\label{norm}
m_1(0)m_2(0)=1,\qquad m_1(0)>0.
\end{gather}
This is a {\it normalization condition}. The properties of the conjugation matrices $[d(z)]^{-\sigma_3}$ listed above
allow to preserve the normalisation condition for all transformations.

The {\it phase function} of our problem is given by
\begin{gather*}
\Phi(z)=\Phi\bigg(z,\frac{n}{t}\bigg)=\frac 1 2 \big(z - z^{-1}\big) + \frac{n}{t}\log z,\qquad
z\in \clos(\C\setminus\R_-).
\end{gather*}
Set $\xi=\frac{n}{t}$ and consider the cross points of the $0$-level lines
for the function $\re\Phi(z,\xi)$, that is, the lines described by
\begin{gather*}
\frac 1 2 \big(z - z^{-1}\big) + \xi\log |z|=0,
\end{gather*}
for different values of $\xi\in \R$.
One of the level lines for all $\xi\in\R$ is evidently the unit circle $|z|=1$. If $\xi\geq 1$ the other two lines are located in the domains $|z|<1$ and $|z|>1$ and are symmetric with respect to the map $z\mapsto z^{-1}$. An elementary analysis shows that the point $-1<z_0(\xi)<0$, where the respective level line crosses the real axis moves monotonously from $-1$ to $0$ when $\xi$ runs the interval $[1, +\infty)$. Thus, the points $z_0(\xi)$ and $z_0^{-1}(\xi)$ meet at point $-1$ for $\xi=1$. The same analysis shows that for $\xi\in [-\infty,0)$ the cross point $z_0(\xi)$ moves monotonously along the interval $[0,1]$ and $z_0(1)=z^{-1}_0(1)=1$. When $\xi\in [-1,1]$, the points $z_0(\xi)$ and $z_0^{-1}(\xi)=\ol{z_0(\xi)}$ lie on the unit circle.

To state the RHP for which $m(z)$ in~\eqref{m_ini} is the unique solution,\footnote{We formulate a RHP which is equivalent to the initial RHPs considered in~\cite{empt19} or~\cite{emt14} in the domain under consideration. Uniqueness is proven in~\cite{emt14}.} we introduce orientations on the jump contour $\Sigma_\delta\cup\Sigma^*_\delta$ according to the symmetry requirements above. The contour $\mathcal C_\delta$ is oriented counterclockwise, $\mathcal C_\delta^*$ is oriented clockwise. On the two symmetric parts
 \begin{gather}\label{defII}
 I:=[q, q_1], \qquad I^*:=\big[q^{-1}, q_1^{-1}\big],
 \end{gather}
the orientation\footnote{In what follows, for $a<b$ the notation $[b,a]$ means that the contour is oriented from $b$ to $a$.} is taken from right to left on $I$ and from left to right on $I^*$. Moreover, all $\mathbb T_{\delta,j}$ and $\mathbb T_{\delta,j}^*$ are supposed to be oriented counterclockwise.
Then $m_+(z)$ (resp.\ $m_-(z))$ will denote the limit from the positive (resp.\ negative) side of the contour. We~assume that these limits exist and $m(z)$ extends to a continuous function on the sides of $\Sigma_\delta\cup\Sigma^*_\delta$ except possibly at the end points of $I$ and $I^*$,
\begin{gather}\label{mathcalJ}
\mathcal J:=\big\{q, q_1, q^{-1}, q_1^{-1}\big\},
\end{gather}
where the square root (not $L^2$-integrable!) singularities are admissible.

Recall that $m_1(z)=m_2\big(z^{-1}\big)=O(z-\ti q)^{-1/2}$ as $z\to \ti q\in\{q, q_1\}$ iff $W(\ti q)=0$, where $W(z)=\ti W(\la(z))$ (cf.~\eqref{wrons}) is the Wronskian of the Jost solutions, $z\in \clos\mathcal D$. If the Wronskian vanishes at $\ti q$, we call $\ti q$ a {\it resonant point}. The general situation is non-resonant, that is, $W(\ti q)\neq 0$. Note that as a function of $z$, the Wronskian takes complex conjugated values on the sides of the contours~\eqref{defII}.

\begin{figure}[ht]\centering
\begin{tikzpicture}

\draw[dotted] (0,0) circle (2.4cm);
\draw[thick] (0,0) circle (2.6cm);
\draw[thick] (0,0) circle (2.2cm);

\draw[thick] (-6.3,0) -- (-3.7,0);
\draw[thick] (-1.7,0) -- (-0.3,0);

\filldraw (-0.3,0) circle (1pt);
\filldraw (-1.7,0) circle (1pt);
\filldraw (-3.7,0) circle (1pt); 
\filldraw (-6.3,0) circle (1pt); 

\draw [->, thick] (3.28,0.2) -- (3.27,0.2);
\draw [->, thick] (0.98,0.2) -- (0.97,0.2);
\draw [->, thick] (1.6,-1.5) -- (1.61,-1.49);
\draw [->, thick] (1.97,-1.69) -- (1.96,-1.7);
\draw [->, thick] (-4.9,0) -- (-4.89,0);
\draw [->, thick] (-0.98,0) -- (-0.99,0);

\draw[thick] (3.3,0) ellipse (0.24cm and 0.2cm);
\filldraw (3.3,0) circle (0.5pt) node[below right]{$\ \ \T_{\delta,j}^*$} node[above right]{$\ \sigma_1v\big(z^{-1}\big)\sigma_1$};
\draw[thick] (1,0) circle (0.2cm) node[below right]{$\ \T_{\delta,j}$} node[above right] {$A_j(z)$};
\filldraw (1,0) circle (0.5pt);

\node at (1.3,-1.3) {$\mathcal C_\delta$};
\node at (2.1,-2.1) {$\mathcal C_\delta^*$};
\node at (-4.9, - 0.3) {$I^*$};
\node at (-5, 0.4) {$\sigma_1v\big(z^{-1}\big)\sigma_1$};
\node at (-0.9, -0.3) {$I$};
\node at (-1, 0.4) {$\begin{psmallmatrix} 1 & 0 \\
	\chi(z) \E^{2t\Phi(z)} & 1 \end{psmallmatrix}$};
\node at (0.2, 1.5) {$\Big(\begin{smallmatrix} 1 & 0 \\
	R (z)\E^{2t\Phi(z)} & 1 \end{smallmatrix}\Big)$};
\node at (2.8, 2) {$\sigma_1v\big(z^{-1}\big)\sigma_1$};
\end{tikzpicture}
\caption{Jump matrix $v(z)$ in Theorem~\ref{RHini}.} 
\end{figure}
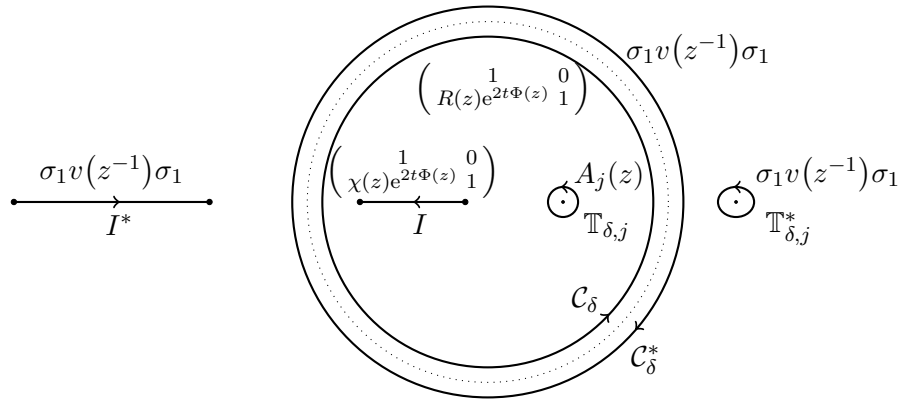

\begin{Theorem} [\cite{empt19, emt14}]
\label{RHini} Let $\delta>0$ be given as in~\eqref{defeps}. For~all $(n,t)\in \Z\times \R_+ $ and $z\in\C$,
 the function~\eqref{m_ini},~\eqref{symcon} is the unique solution of the following RH problem:
 to find a vector-function $m(z)=(m_1(z), m_2(z))$ holomorphic in $\C\setminus(\Sigma_\delta\cup\Sigma^*_\delta)$ which
 \begin{itemize}
 \item is continuous up to the boundary, except of possibly points in $\mathcal J$~\eqref{mathcalJ},
 \item satisfies conditions~\eqref{symcon},~\eqref{norm} and the jump condition
\begin{gather}\label{minin}
m_{+}(z)=m_{-}(z) v(z), \qquad z\in\Sigma_\delta\cup\Sigma^*_\delta,
\\
\label{vini} v(z) = \begin{cases}
\begin{pmatrix}
1 & 0 \\ \chi(z) \E^{2t\Phi(z)} & 1
\end{pmatrix}\!,
& z \in I,
\\[2.5ex]
\begin{pmatrix}
1&0\\ R(z) \E^{2t\Phi(z)} & 1
\end{pmatrix}\!,
& z \in \mathcal C_\delta,
\\[1ex]
A_j(z), & z\in\mathbb T_{\delta, j},\ z_j\in\si_{\rm d},
\\
\si_1 v(z^{-1})\si_1, & z \in \Sigma^*_\delta,
\end{cases}
 \end{gather}
where $\Phi(z)=\Phi(z,\frac{n}{t})$ is the phase function, matrices $A_j(z)$ are defined by~\eqref{A_j} and~$R(z)$ is the holomorphic continuation of the initial right reflection coefficient~\eqref{simpl3}. The function~$\chi(z)$ is defined by
\begin{gather}\label{defchi}
\chi(z):=\frac{\big(z - z^{-1}\big)\big[\zeta^{-1} - \zeta\big](z-\I 0)}{| W(z)|^2},\qquad z\in I,
\end{gather}
with $\zeta(z)$, $z\in \ol{\mathcal D}$, connected with $z$ by the Joukovski map
 $\la=b+a(\zeta + \zeta^{-1})$. Here $W(z)=\ti W(\la(z))$ is the Wronskian of the Jost solutions defined in~\eqref{wrons},
\item in vicinities of the points in $\mathcal J$, $m(z)$ has the following behavior:
\begin{itemize}
\item if $\ti q\in \{q, q_1\}$ is nonresonant, then $m(z)=(O(1), O(1))$ as $z\to \ti q$, $z\to\ti q^{-1}$,
\item if $\ti q\in \{q, q_1\}$ is resonant, i.e., if $\chi(z)=O\big(\frac{1}{\sqrt{z-\ti q}}\big)$ as $z\to \ti q$, then
\begin{alignat*}{3}
& m(z)= \bigg(O\bigg(\frac{1}{\sqrt{z-\ti q}}\bigg), O(1)\bigg), \qquad&&
z\to \ti q,& \\
& m(z)= \bigg(O(1), O\bigg(\frac{1}{\sqrt{z-\ti q^{-1}}}\bigg)\bigg),\qquad && z\to \ti q^{-1}.&
\end{alignat*}
\end{itemize}
\end{itemize}
 \end{Theorem}

\begin{Remark}\label{continu}
According to~\eqref{decay}, $\ol{W(z)}$ admits an analytic continuation in a small vicinity of the interval $I$ (see~\cite[equation~(2.11)] {empt19}). Respectively, in this vicinity there exists an analytic continuation $X(z)$ of~$\chi(z)$ such that $X_\pm(z)=\pm \I|\chi(z)|$ for $z\in I$. The function $X(z)$ does not have other jumps in this vicinity.
 \end{Remark}

Note that in the VDO region, the off-diagonal matrix elements of the jump matrix $v(z)$ grow exponentially with respect to $t$ for $z\in I\cup I^*$. The same is true for $v(z)$ on those contours $\mathbb T_{\delta, j}\cup \mathbb T_{\delta, j}^*$ which correspond to the origins $0>z_j>z_0(\xi)$. The remaining parts of the jump matrices are asymptotically close to the identity matrix as $t\to \infty$. In the next sections we perform a series of conjugation/deformation steps which transform the initial RHP of Theorem~\ref{RHini} to the equivalent problem with a jump matrix which is asymptotically close as $t\to\infty$ to a piecewise constant matrix with respect to $z$. This limiting matrix also depends on $\xi$ as a piecewise constant matrix, and the respective (so called model) RHP has
a unique solution which can be found explicitly in terms of the Riemann theta-function. Let us emphasize that for the VDO region we propose transformations which lead to the absence of any additional parametrix problems. The first transformation of the initial problem solution is associated with the so called $g$-function method first introduced for the KdV equation in~\cite{dvz}. In our case, this $g$-function is a normalized Abel integral associated with the two-sheeted Riemann surface glued via the cuts along the continuous spectrum $\mathfrak S=[b-2a, b+2a]\cup [-1, 1]$. In fact, the $g$-function is
a linear combination (dependent on~$\xi$) of the normalized Abel differentials of the second and third type which are involved in~the exponential part of the Baker--Akhiezer function corresponding to the finite gap solution of~the Toda lattice associated with the two-band spectrum $\mathfrak S$. It~is crucial for our endeavor to understand in detail the properties of the $g$-function as a function of the spectral parameter $\la$. The next section is devoted to this subject.

\section[g-function as an Abel integral and its connection with the Baker--Akhiezer function]
{$\boldsymbol g$-function as an Abel integral and its connection\\ with the Baker--Akhiezer function}\label{sec3}

Let $\mathbb M$ be the two-sheeted Riemann surface associated with $\mathfrak S$, i.e., with the function
\begin{gather*}
\mathcal R^{1/2}(\la)=-\sqrt{\big(\la^2 - 1\big)\big((\la - b)^2 - 4 a^2\big)},
\end{gather*}
with sheets $\Pi_U$ and $\Pi_L$ glued along the cuts over the intervals $[b-2a, b+2a]$ and $[-1, 1]$. The~indices $U$ and $L$ label the upper and lower sheets of the surface. We~denote by $p=(\la, \pm)$ the points of $\mathbb M$, with $(\infty, \pm):=\infty_\pm$; and $p^*=(\la,\mp)$ for $p=(\la, \pm)$ denotes the sheet exchange map. Choose a canonical basis of $\mathfrak a$ and $\mathfrak b$ cycles on $\mathbb M$ as follows: the cycle $\mathfrak b$ surrounds the interval $[b-2a, b+2a]$ counterclockwise on $\Pi_U$ and the cycle $\mathfrak a$ passes from $b+2a$ to $-1$ on the upper sheet and back on the lower sheet. The part of this cycle on $\Pi_U$ we denote by $J^U$ and treat it as a contour on $\mathbb M$. Its projection lies on the interval $[b+2a, -1]$ which we call the gap. The lower part of $\mathfrak a$ has the same projection on the gap, and is considered as the contour $J^L$ which passes from $-1$ to $b+2a$ on $\Pi_L$.

Let $\Omega_0$ be the Abel differential of the second kind on $\mathbb M$ with second order poles at $\infty_+$ and~$\infty_-$ and~let
$\omega_{\infty_+, \infty_-}$ be the Abel differential of the third kind with logarithmic poles at $\infty_+$ and $\infty_-$, both normalized as
\begin{gather*}
\int_{\mathfrak a} \Omega_0=\int_{\mathfrak a} \omega_{\infty_+,\infty_-}=0.
\end{gather*}
As it is known,
\begin{gather}\label{struct}
\Omega_0=\frac{(\la - \nu_1)(\la - \nu_2)}{\mathcal R^{1/2}(\la)}\,{\rm d}\la, \qquad \omega_{\infty_+,\infty_-}=\frac{\la - \nu_3}{\mathcal R^{1/2}(\la)}\,{\rm d}\la,
\end{gather}
where $\nu_i\in\R$ for $i=1,2,3$. Moreover, $\nu_3\in (b+2a, -1)$, and at least one of the points $\nu_1$ or $\nu_2$ also lies in the gap $(b+2a, -1)$. Consider the Abel integral given by
\begin{gather*}g(p,\xi):=\int_1^p \Omega_0 +\xi \int_1^p \omega_{\infty_+, \infty_-},\end{gather*}
where $\xi\in\R$ is a parameter.
On $\Pi_U$ we denote it by $g(\la, \xi)$, that is,
\begin{gather}\label{defgee}
g(\la, \xi)=\int_1^\la \frac{(\la -\nu_1)(\la - \nu_2)-\xi(\la - \nu_3)}{\mathcal R^{1/2}(\la)}\, {\rm d}\la= \int_1^\la \frac{(\la -\mu_1(\xi))(\la - \mu_2(\xi))}{\mathcal R^{1/2}(\la)}\, {\rm d}\la.
\end{gather}
Since
\begin{gather}\label{normper}
\int_{b+2a}^{-1} \frac{(\la -\mu_1(\xi))(\la - \mu_2(\xi))}{\mathcal R^{1/2}(\la)}\, {\rm d}\la=0,
\end{gather}
then $\mu_i(\xi)\in\R$, these points do not coincide, and at least one of them belongs to the gap.
By~definition of $\Omega_0$ we have
 \begin{gather*}
 \frac{(\la -\nu_1)(\la - \nu_2)}{\mathcal R^{1/2}(\la)}= -1 +O\big(\la^{-2}\big),
 \end{gather*}
 that is,
\begin{gather*}
\frac{(\la -\mu_1(\xi))(\la - \mu_2(\xi))}{\mathcal R^{1/2}(\la)}+1= \frac{-\mu_1(\xi) - \mu_2(\xi) + b}{\la} + O\big(\la^{-2}\big)=\frac{\xi}{\la} +O\big(\la^{-2}\big),
\end{gather*}
which implies
\begin{gather}\label{defmu2}
\mu_1(\xi)=b-\xi - \mu_2(\xi).
\end{gather}
By~\eqref{normper},
\begin{gather*}
\int_{b+2a}^{-1}\frac{(\la -\mu_i(\xi))(\la - b+\xi +\mu_i(\xi)))}{\mathcal R^{1/2}(\la)}\, {\rm d}\la=0,
\end{gather*}
that is, $\mu_i(\xi)$ are the zeros of the quadratic equation
\begin{gather*}
\mu^2\int_{\mathfrak a} \frac{{\rm d}\la}{\mathcal R^{1/2}(\la)} + \mu (\xi - b) \int_{\mathfrak a} \frac{{\rm d}\la}{\mathcal R^{1/2}(\la)}-\int_{\mathfrak a} \frac{\la^2 + \la(\xi - b)}{\mathcal R^{1/2}(\la)}\,{\rm d}\la=0.
\end{gather*}
With the notations
\begin{gather}\label{gamma12}
\Gamma_1=\frac{\int_{\mathfrak a} \frac{\la^2{\rm d}\la}{\mathcal R^{1/2}(\la)}}{\int_{\mathfrak a} \frac{{\rm d}\la}{\mathcal R^{1/2}(\la)}},\qquad
\Gamma_2=\frac{\int_{\mathfrak a} \frac{\la\, {\rm d}\la}{\mathcal R^{1/2}(\la)}}{\int_{\mathfrak a} \frac{{\rm d}\la}{\mathcal R^{1/2}(\la)}},
\end{gather}
we infer
\begin{gather}\label{raz}
\mu_{1,2}(\xi)=\frac{1}{2}\Big( b-\xi \pm \sqrt{(b-\xi)^2 + 4(\Gamma_1 + (\xi - b)\Gamma_2)}\Big).
\end{gather}

\begin{Lemma} The functions $\mu_i(\xi)$, $i=1,2$, are monotonically decreasing with respect to $\xi\in\R$. For~$\xi\in (\xi_{\aleph +1}, \xi_0)$,\footnote{In~\cite{emt14} and~\cite{empt19}, these values were denoted by $\xi_{{\rm cr}, 1}^\prime $ and $\xi_{{\rm cr}}^\prime$.} where
\begin{gather}\label{bound}
\xi_{\aleph+1}=b+\frac{1 - \Gamma_1}{1+\Gamma_2}, \qquad \xi_0= b +\frac{\Gamma_1 - (b+2a)^2}{b+2a - \Gamma_2},
\end{gather}
we have
$\mu_i(\xi)\in (b+2a, -1)$, $i=1,2$.
\end{Lemma}
\begin{proof}
Differentiating~\eqref{raz} with respect to $\xi$ implies
\begin{gather*}
2\frac{\rm d}{{\rm d}\xi}\mu_i(\xi)= -1\pm \frac{\xi - b +2\Gamma_2}{\sqrt{(b-\xi)^2 + 4(\Gamma_1 + (\xi - b)\Gamma_2)}}.
\end{gather*}
We observe that
\begin{gather*}
\Gamma_1>1, \qquad \Gamma_2<-1,\qquad |\Gamma_1|>|\Gamma_2|.
\end{gather*}
Inequality
\begin{gather*}
|\xi - b +2\Gamma_2|<\sqrt{(b-\xi)^2 + 4(\Gamma_1 + (\xi - b)\Gamma_2)}
\end{gather*}
holds if $\Gamma_2^2<\Gamma_1$. The last one follows from the Cauchy inequality
\begin{gather*}
\bigg|\int_{\mathfrak a} \frac{\la \,{\rm d}\la}{\mathcal R^{1/2}(\la)}\bigg|<\sqrt{
\int_{\mathfrak a} \frac{\la^2 {\rm d}\la}{\mathcal R^{1/2}(\la)}}\ \sqrt{
\int_{\mathfrak a} \frac{ {\rm d}\la}{\mathcal R^{1/2}(\la)}}.
\end{gather*}
Thus, $\mu_i(\xi)$ are monotonically decreasing with respect to $\xi$. Assume that $\mu_1(\xi)<\mu_2(\xi)$. A~trivial analysis shows that the value $\xi_0$ corresponds to the location $\mu_1(\xi_0)=b+2a$, that is,
\begin{gather*}
2(b+2a)=b-\xi_0- \sqrt{(b-\xi_0)^2 + 4(\Gamma_1 + (\xi_0 - b)\Gamma_2)}.
\end{gather*}
This implies the second equation in~\eqref{bound}. The location $\mu_2(\xi_{\aleph+1})=-1$ provides the first formula in~\eqref{bound}.
From~\eqref{defmu2} it follows that
\begin{gather*}
\xi_0=-2a -\mu_1(\xi_0),\qquad
\xi_{\aleph+1}=b+1 -\mu_{2}(\xi_{\aleph+1}),
\end{gather*}
that is,
\begin{gather*}
\xi_0 - \xi_{\aleph+1}=|b+2a| - 1 + \mu_{2}(\xi_{\aleph+1}) -\mu_1(\xi_0)>0,
\end{gather*}
since $\mu_1(\xi_0),\ \mu_{2}(\xi_{\aleph+1})\in (b+2a, -1)$.
 \end{proof}

 Let $\varepsilon>0$ be an arbitrary small number and let $\xi_0$ and $\xi_{\aleph+1}$ be defined by~\eqref{bound} and~\eqref{gamma12}. For~any $\xi\in [\xi_{\aleph+1} + \varepsilon, \xi_0 - \varepsilon]$,
 both points $\mu_1(\xi)$ and $\mu_2(\xi)$ are inner points of the gap.
The set of level lines $\re g=0$ consists of the two intervals $[b-2a, b+2a]$ and $[-1, 1]$ and an infinite
contour which intersects the real axis at $\mu_0(\xi)$ such that
\begin{gather}\label{dispoz}
\mu_1(\xi)<\mu_0(\xi)<\mu_2(\xi).
\end{gather}
\begin{Lemma} \label{discre}
The real-valued function $\mu_0(\xi)$ implicitly given by $\re g(\mu_0(\xi), \xi)=0$ is monotonic with $\frac{\rm d}{{\rm d}\xi}\mu_0(\xi)<0$ for $\xi\in (\xi_{\aleph+1}, \xi_0)$.
Moreover,
\begin{gather}\label{lim}
\lim_{\xi\to \xi_0} \mu_0(\xi)= \lim_{\xi\to \xi_0} \mu_1(\xi)=b+2a,\qquad
\lim_{\xi\to \xi_{\aleph+1}} \mu_0(\xi)= \lim_{\xi\to \xi_{\aleph+1}} \mu_2(\xi)=-1.
\end{gather}
\end{Lemma}

\begin{proof} By~\eqref{raz} we have $\mu_1(\xi)\mu_2(\xi)=(b-\xi)\Gamma_2 - \Gamma_1$. This implies with~\eqref{defmu2} that
$\mu_0(\xi)$ is given implicitly by
\begin{gather*}
\int^{-1}_{\mu_0(\xi)}\frac{(\la - \mu_1(\xi))(\la - \mu_2(\xi))}{\mathcal R^{1/2}(\la)}\,{\rm d}\la=\int^{-1}_{\mu_0(\xi)}\frac{\la^2 +(\xi - b)\la +((b-\xi)\Gamma_2 - \Gamma_1)}{\mathcal R^{1/2}(\la)}\,{\rm d}\la=0.
\end{gather*}
Differentiating with respect to $\xi$ implies
\begin{gather*}
\frac{\rm d}{{\rm d}\xi}\mu_0=\int^{-1}_{\mu_0}\frac{\la - \Gamma_2}{\mathcal R^{1/2}(\la)}\, {\rm d} \la\,\frac{R^{1/2}(\mu_0)}{(\mu_0 - \mu_1)(\mu_0 - \mu_2)}.
\end{gather*}
Since the second multiplier is negative,
it is sufficient to prove that the integral is positive. But
\begin{gather*}
\int^{-1}_{\mu_0}\frac{\la - \Gamma_2}{\mathcal R^{1/2}(\la)}\,{\rm d}\la = \int^{-1}_{\mu_0}\frac{\la}{\mathcal R^{1/2}(\la)}\,{\rm d}\la - \int^{-1}_{\mu_0}\frac{{\rm d}\la}{\mathcal R^{1/2}(\la)} \int^{-1}_{b+2a}\frac{\la}{\mathcal R^{1/2}(\la)}\,{\rm d}\la
\bigg(\int^{-1}_{b+2a}\frac{{\rm d}\la}{\mathcal R^{1/2}(\la)}\bigg)^{-1}\!\!,
\end{gather*}
that is, we have to prove that
\begin{gather*}
\frac{\int^{-1}_{\mu_0}\frac{\la\,{\rm d}\la}{\mathcal R^{1/2}(\la)} }{\int^{-1}_{\mu_0}\frac{{\rm d}\la}{\mathcal R^{1/2}(\la)}} > \frac{ \int^{-1}_{b+2a}\frac{\la\,{\rm d}\la}{\mathcal R^{1/2}(\la)}}{
\int^{-1}_{b+2a}\frac{{\rm d}\la}{\mathcal R^{1/2}(\la)}}.
\end{gather*}
This inequality is true by the mean value theorem.
Equalities~\eqref{lim} were proven in~\cite{emt14}.
\end{proof}

Let us recall the Baker--Akhiezer function for a finite gap solution $\big\{\hat a(n,t), \hat b(n,t)\big\}$ of the Toda lattice equation associated with the spectrum
$\mathfrak{S}=[b-2a, b+2a]\cup [-1, 1]$ and the initial Dirichlet divisor $p_0=(\la(0,0),\si(0,0))$, $\si(0,0)\in\{+, -\}$.
In our case the divisor consists of one point on the Riemann surface $\mathbb M$ with projection on the closed gap of the spectrum and it will later depend on the slow variable $\xi$.

Let $\zeta$ be the holomorphic Abel differential on $\mathbb M$ normalized as $\int_{\mathfrak a}\zeta=1$ and let $\int_{\mathfrak b}\zeta=:\tau\in\I \R_+$
be its $\mathfrak b$-period. Introduce the Abel map $A(p):=\int_{b-2a}^p \zeta$. It~is an odd function on~$\mathbb M$, $A(p^*)=-A(p)$.
Moreover, it has a jump along the $\mathfrak a$-cycle, which we interpret as a union $J^U\cup J^L$. Then $A_+(p)-A_-(p)=-\tau$ as $p\in J^U\cup J^L$. We~set $A(b+2a):=A_+(b+2a)=-\tau/2$ and $A(p_0)=A_+(p_0)$, where $p_0\in J^U\cup J^L$ is the initial Dirichlet divisor.

Let $\Xi=\frac{\tau}{2} + \frac 1 2$ be the Riemann constant and
\begin{gather}\label{defcycle}
\Lambda=\int_{\mathfrak b}\omega_{\infty_+, \infty_-}\in \I \R,\qquad
U=\int_{\mathfrak b}\Omega_0\in \I\R,
\end{gather}
be the $\mathfrak b$-periods of the Abel differentials~\eqref{struct}.
Following~\cite{tjac}, we introduce the notations
\begin{gather*}
Z(p,n,t):=A(p) - A(p_0) -n\frac{\Lambda}{2\pi\I} - t\frac{U}{2\pi\I}-\Xi,
\\
Z(n,t):=Z(\infty_+, n,t).
\end{gather*}
Evidently, $\theta (Z(p,0,0))=0$ iff $p=p_0$, where
\begin{gather*}
\theta(v):=\theta(v\,|\,\tau)=\sum_{m\in\Z} \exp\big(\pi\I m^2\tau + 2\pi\I m v\big)
\end{gather*}
is the Jacobi theta function. Recall that the time-dependent Baker--Akhiezer function for the finite gap Toda lattice solution with spectral data as above has the form~\cite{tjac}
\begin{gather*}
\Psi(p,n,t)= C(n,t)\frac{\theta(Z(p,n,t))}{\theta(Z(p,0,0))}\exp\bigg(n \int_1^p\omega_{\infty_+, \infty_-} + t \int_1^p\Omega_0\bigg)
\\ \hphantom{\Psi(p,n,t)}
{}=C(n,t)\frac{\theta(Z(p,n,t))}{\theta(Z(p,0,0))}\exp(t g(p,\xi)),\qquad \xi=\frac{n}{t}.
\end{gather*}

 Here
$C(n,t)$ is a positive constant (with respect to $p$) which provides the equalities
\begin{gather*}
\lim_{p\to\infty_\pm} \Psi(p^*,n,t)\Psi(p,n,t)=1, \qquad \Psi(p,0,0)=1,
\end{gather*}
and
\begin{gather*}
\frac{C(n+1,t)}{C(n,t)}=\sqrt{\frac{\theta(Z(n-1,t))}{\theta(Z(n+1,t))}}>0.
\end{gather*}
As is known, for each $n$ and $t$ fixed, the Baker--Akhiezer function is a meromorphic
function of~$p$ on $\mathbb M$ with a simple pole at $p_0$. Respectively,
the vector function $(\Psi(p^*, n, t), \Psi(p,n,t))$ does not have jumps on $\mathbb M$, and the vector function
\begin{gather*}
\hat m(p):= \bigg(\Psi(p^*, n, t)\exp\bigg(\!{-}t g\bigg(p^*,\frac{n}{t}\bigg)\!\bigg), \Psi(p,n,t)\exp\bigg(\!{-}t g\bigg(p,\frac{n}{t}\bigg)\!\bigg)\bigg),
\end{gather*}
has an evident jump along $\mathfrak a$,
\begin{gather*}
\hat m_+(p)=\hat m_-(p)\E^{-(n \Lambda + t U)\sigma_3},\qquad
p\in J^U\cup J^L.
\end{gather*}
Here we took into account that for $p\in J^U\cup J^L$,
\begin{gather*}
\bigg[\int_1^p \omega_{\infty_+, \infty_-}\bigg]_+
-\bigg[\int_1^p \omega_{\infty_+, \infty_-}\bigg]_-=-\Lambda,\qquad
\bigg[\int_1^p\Omega_0\bigg]_+ - \bigg[\int_1^p\Omega_0\bigg]_-=-U.
\end{gather*}
Note that since $g(p^*,\xi)=-g(p,\xi)$,
\begin{gather*}
\hat m_1(\infty_+)\hat m_2(\infty_+)=1,\qquad
\hat m(p)=\hat m(p^*)\sigma_1.
\end{gather*}
The function
\begin{gather*}
f(p):=f_\infty \frac{\theta(Z(p,0,0))}{\theta\big(A(p) - \frac{1}{2}\big)}, \qquad f_\infty:=\sqrt{\frac{\theta\big(A(\infty_+)-\frac 1 2\big)
\theta\big(A(\infty_-)-\frac 1 2\big)}{\theta(Z(\infty_+,0,0))\theta(Z(\infty_-,0,0))}},
\end{gather*}
has a simple pole at the branch point $b+2a$ and a simple zero at $p_0$. Moreover,
\begin{gather*}
f(\infty_+)f(\infty_-)=1, \qquad
f_+(p^*)=f_-(p^*)\E^{-\I\Delta}, \qquad
f_+(p)=f_-(p)\E^{\I\Delta}, \qquad p\in J^U,
\end{gather*}
where
\begin{gather}\label{defdelta}
\Delta=-2\pi \bigg(A(p_0) +\frac{\tau}{2}\bigg)\in\R.
\end{gather}
Note that~\eqref{defdelta} can be rewritten as the Jacobi inversion problem
\begin{gather}\label{jacinv}
\int_{b+2a}^{p_0}\zeta=-\frac{\Delta}{2\pi}\quad (\text{mod}\, 1)
\end{gather}
and allows us to compute uniquely the divisor point $p_0$ for any given real valued $\Delta$.
Summing up the considerations above, we proved the following
\begin{Lemma}\label{lem:fg} The vector function
\begin{align*}
\ti m(p)&=\big(\ti m_1(p,n,t),\ti m_2(p,n,t)\big) \nonumber
\\
& =\bigg(\Psi(p^*, n, t)f(p^*)\exp\bigg({-}t g\bigg(p^*,\frac{n}{t}\bigg)\bigg), \Psi(p,n,t)f(p)\exp\bigg({-}t g\bigg(p,\frac{n}{t}\bigg)\bigg)\bigg)\nonumber
\\
& = \ti C(n,t)\bigg(\frac{\theta(Z(p^*,n,t))}{\theta\big(A(p^*)-\frac{1}{2}\big)},
\frac{\theta(Z(p,n,t))}{\theta\big(A(p)-\frac{1}{2}\big)} \bigg),\qquad \text{where}\quad
\ti C(n,t):=C(n,t)f_\infty,
\end{align*}
 solves the following RHP on $\mathbb M{:}$ to find a holomorphic vector-function $\ti m(p)$ on $\mathbb M\setminus\big(J^U\cup J^L\big)$, which satisfies
\begin{itemize}
\item the jump condition
\begin{gather}\label{jumpmod}
\ti m_+(p)=\ti m_-(p)\E^{-(n\Lambda + t U +\I\Delta)\sigma_3},
\end{gather}
\item the symmetry condition
\begin{gather}\label{tim}
\ti m(p^*)=\ti m(p)\sigma_1 \qquad \text{for}\quad p\in \mathbb M\setminus \big(J^U\cup J^L\big),
\end{gather}
\item the normalization condition $\ti m_1(\infty_+)\ti m_2(\infty_+)=1$.
\item Both components of $\ti m(p)$ have simple poles at the branch point $p=b+2a$ and no other singularities.
\end{itemize}
\end{Lemma}
Note that the constant $\ti C(n,t)$ in Lemma \ref{lem:fg} satisfies
\begin{gather*}
\frac{\ti C(n+1,t)}{\ti C(n,t)}=\sqrt{\frac{\theta(Z(n-1,t))}{\theta(Z(n+1,t))}}>0.
\end{gather*}
Together with Theorem 9.48 of~\cite{tjac} it implies
\begin{Corollary}\label{cor:fgas}
For the vector function $\ti m(p)=\ti m(p,n,t)$ the following holds
\begin{gather*}
\frac{\ti m_1(\infty_+, n, t)}{\ti m_1(\infty_+, n+1, t)}=
\frac{\sqrt{\theta(Z(n-1,t))\theta(Z(n+1,t))}}{\theta(Z(n,t))}=\frac{\hat a(n,t)}{\ti a},\end{gather*}
where $\ti a=\mathrm{Cap}\,\mathfrak S$ is the logarithmic capacity of the spectrum $\mathfrak S$.
\end{Corollary}
Introduce the product
\begin{gather*}
\ti h(p):=\ti m_1(p)\ti m_2(p)= \ti C^2(n,t)\frac{\theta(Z(p^*,n,t))}{\theta\big(A(p^*)-\frac{1}{2}\big)}
\frac{\theta(Z(p,n,t))}{\theta\big(A(p)-\frac{1}{2}\big)},
\end{gather*}
and let $ h(\la):=\ti h(p)$ for $p=(\la, +)$.
The function $\ti h(p)$ has a double pole on $\mathbb M$ at the branch point $b+2a$, that is, $h(\la) $ has a simple pole at $b+2a$. Moreover, $\theta(Z(p,n,t))$ has the only zero at $p(n,t)=(\la(n,t), \pm)\in J^U\cup J^L$, which is the unique solution of the Jacobi inversion problem
\begin{gather}\label{solJ}
\int_{p_0}^{p(n,t)}\zeta=n\frac{\Lambda}{2\pi \I} + t\frac{U}{2\pi \I},
\end{gather}
$\theta(Z(p^*,n,t))$ has a simple zero at the involution point $p^*(n,t)$. Thus $h(\la(n,t))=0$, and it is a~simple zero of $h$. We~observe that from the jump and symmetry conditions it follows that $ \ti h(p)$ does not have jumps on $\mathbb M$, moreover, $\ti h(p)=\ti h(p^*)$, $p\in\mathbb M$. This means that $h(\la)$ does not have jumps along the spectrum
$\mathfrak S$ and on the gap $[b+2a, -1]$.
The normalisation condition implies $\lim_{\la\to\infty} h(\la)=1$. Hence $h(\la)$ is a meromorphic function on $\C$, i.e.,
$h(\la)=\frac{\la - \la(n,t)}{\la - b-2a}$.
\begin{Corollary}
Let $\la(n,t)\in [b+2a, -1]$ be the projection on $\C$ of the Dirichlet eigenvalue $p(n,t)$ given by~\eqref{solJ}. Then
\begin{gather*}
\lim_{p\to\infty_+}p \left(\ti m_1(p)\ti m_2(p)-1\right)=b+2a -\la(n,t).
\end{gather*}
\end{Corollary}
We recall that the trace formula in our case looks like
\begin{gather*}
\hat b(n,t)=\frac{1}{2}\left(1+b-2a + b+2a -1-2\la(n,t)\right)=b-\la(n,t).
\end{gather*}
Therefore,
\begin{gather*}
\lim_{p\to\infty_+}p \left(\ti m_1(p)\ti m_2(p)-1\right)=\hat b(n,t)+2a.
\end{gather*}
Since the problem~\eqref{jacinv} has a unique solution $p_0$ for any real $\Delta$, we can treat $\Delta$ as the ini\-tial data to choose the representative $\big\{\hat a(n,t), \hat b(n,t)\big\}$ for the isospectral set of finite gap potentials with spectrum $\mathfrak S$. To emphasise this dependence we denote the representative as $\big\{\hat a(n,t,\Delta), \hat b(n,t,\Delta)\big\}$. In turn, the solution of the RHP with jump~\eqref{jumpmod}
we denote as $\ti m(p,n,t,\Delta)$. We~proved the following
\begin{Theorem}\label{theorRHs}
Let $\ti m(p,n,t,\Delta)$ be the unique solution of the RHP in Lemma~$\ref{lem:fg}$. Then
\begin{gather*}
\lim_{p\to\infty_+}\ti a \frac{\ti m_1(p, n, t,\Delta)}{\ti m_1(p, n+1, t,\Delta)}=
\hat a(n,t,\Delta),
\\
 \lim_{p\to\infty_+}p \left(\ti m_1(p,n,t,\Delta)\ti m_2(p,n,t,\Delta)-1\right)-2a=\hat b(n,t,\Delta),
\end{gather*}
where $\big\{\hat a(n,t,\Delta), \hat b(n,t,\Delta)\big\}$ is the finite gap solution of~\eqref{tl1} with two band spectrum $\mathfrak S=[b-2a, b+2a]\cup [-1,1]$ and initial Dirichlet divisor $p_0$ given by the Jacobi inversion~\eqref{jacinv}.
\end{Theorem}
\begin{Remark}\label{remdef}
For convenience of the reader, we recall from~\cite{tjac} that the finite gap solution
$\{\hat a(n,t,\Delta), \hat b(n,t,\Delta)\}$ corresponding to the initial phase $\Delta$
and spectrum $\mathfrak S$ is given by
\begin{gather}
\hat a(n,t,\Delta)=\ti a \frac{\sqrt{\theta\big(\frac{(n-1)\Lambda}{2\pi\I} +\frac{t U}{2\pi\I}-\frac{\Delta}{2\pi}+A \big) \theta\big(\frac{(n+1)\Lambda}{2\pi\I}+\frac{t U}{2\pi\I}-\frac{\Delta}{2\pi}+A\big)}}{\theta\big(\frac{n\Lambda}{2\pi\I}+\frac{t U}{2\pi\I}-\frac{\Delta}{2\pi}+A \big)},\nonumber
\\
\hat b(n,t,\Delta) =\ti b +\frac{1}{Y}\frac{\pa}{\pa w}\log \frac{\theta\big(\frac{(n-1)\Lambda}{2\pi\I} +\frac{t U}{2\pi\I}-\frac{\Delta}{2\pi}+A+w \big)}{\theta\big(\frac{n\Lambda}{2\pi\I} +\frac{t U}{2\pi\I}-\frac{\Delta}{2\pi}+A+w \big)},
\label{def ab}
\end{gather}
where
\begin{gather*}
A:=\frac{1}{2} +\int_{\infty_+}^{b-2a}\zeta,\qquad
Y=\int_{\mathfrak a}\frac{{\rm d}\la}{\mathcal R^{1/2}(\la)},\qquad
\ti b=b-\frac{1}{Y}\int_{\mathfrak a} \frac{\la\, {\rm d}\la}{ R^{1/2}(\la)},
\end{gather*}
$\zeta$ is the normalized holomorphic Abel differential, $\ti a=\mathrm{Cap}\,\mathfrak S$, and $U$, $\Lambda$ are defined by~\eqref{defcycle}.
\end{Remark}

\section{Reduction of the initial RHP to the model RHP}\label{Sec:4}

From now we work again in the variable $z$. Let us identify the upper sheet $\Pi_U$ of the Riemann surface $\mathbb M$ with the domain~\eqref{defmathcald} and the lower sheet with~\eqref{defdstar}. The image of $J^U$ under the map $p\mapsto z$ we denote by $J$, preserving the orientation, i.e., $J=[q_1, -1]$ is oriented from right to left. As for the image $J^*$ of $J^L$, we change its orientation in accordance with our symmetry requirements, i.e., $J^*=\big[ q_1^{-1}, -1\big]$ is oriented from left to right. The other contours used here are already defined by~\eqref{defII},~\eqref{Ceps},~\eqref{circles},~\eqref{sig}. In this section, we perform two transformations (steps) which transform the solution of the initial RHP (Theorem \ref{RHini}) to the solution of the RHP with the jump matrix which is close as $t\to \infty$ to a piecewise constant jump matrix everywhere on the jump contour, without exceptional points (parametrices).

{\it Step 1}. Set
\begin{gather*}
y_k=y_k(\xi)=z(\mu_k(\xi))\in J\qquad \text{for}\quad k=0, 1,2,
\end{gather*}
where $\mu_k(\xi)$ are defined in Section~\ref{sec3}. From~\eqref{dispoz} we have
$y_2(\xi)<y_0(\xi)<y_1(\xi)$.
With these definitions at hand, the $g(p)$-function~\eqref{defgee} is given in terms of $z$ by
\begin{gather*}
g(z, \xi) = \frac12 \int_1^z \frac{(s-y_1)\big(s-y_1^{-1}\big)(s-y_2)\big(s-y_2^{-1}\big)}
{\sqrt{(s-q_1)\big(s-q_1^{-1}\big)(s-q)\big(s-q^{-1}\big)}} \frac{{\rm d}s}{s^2}, \qquad
z\in \C\setminus (-\infty, 1).
\end{gather*}
Thus, the level lines $\re g(z,\xi)=0$ which are different from the unit circle $\mathbb T$ and intervals $I$ and $I^*$, cross the real axis at the points $y_0(\xi)$ and $y_0^{-1}(\xi)$.
Similar to~\cite[Lemma 3.1]{empt19} and \cite[Lemma 5.3]{emt14} we establish the following

\begin{figure}[ht]
\centering
\begin{tikzpicture}

\clip (-4,-4) rectangle (8.5,0);

\fill[gray!5] (4.5,-2) circle (2.9cm);
\path [fill=white] [out=120, in=110] (2.2,-2) to (4.5,-2) [out=240, in=250] (2.2,-2) to (4.5,-2);
\draw[thick] (4.5,-2) circle (2.9cm);
\path [fill=gray!5] [out=60, in=340] (-0.5,-2) to (-3.5,0.5) -- (-3.5,-4.5) [out=300, in=20] (-0.5,-2) to (-3.5,-4.5);

\draw[thick] (3,-2) -- (3.8,-2);
\draw[thick] (-3.2,-2) -- (-1.4,-2);

\draw (-1,-2) ellipse (0.12cm and 0.1cm);
\filldraw (-1,-2) circle (0.5pt) node[above]{\small{${\T_{\delta, j}^*}$}};
\draw (2.7,-2) circle (0.1cm);
\filldraw (2.7,-2) circle (0.5pt) node[above]{\small{${\T_{\delta, j}}$}};
\draw[thin] (1.9,-2) circle (0.1cm);
\filldraw (1.9,-2) circle (0.5pt); 
\draw (0.5,-2) ellipse (0.12cm and 0.1cm);
\filldraw (0.5,-2) circle (0.5pt);

\draw[out=60, in=340, gray, dotted] (-0.5,-2) to (-3.5,0.5);
\draw[out=300, in=20, gray, dotted] (-0.5,-2) to (-3.5,-4.5);
\draw[out=120, in=110, gray, dotted] (2.2,-2) to (4.5,-2);
\draw[out=240, in=250, gray, dotted] (2.2,-2) to (4.5,-2);

\node at (-2.6,-0.7) {$\re g < 0$};
\node at (3.5,-0.7) {$\re g < 0$};
\node at (0.5,-0.7) {$\re g > 0$};
\node at (6.7,-0.6) {$\T$};
\filldraw (3.8,-2) circle (1pt) node[below]{\small{$q$}};
\filldraw (3,-2) circle (1pt) node[below]{\small{$q_1$}};
\filldraw (4.5,-2) circle (1pt) node[below]{${0}$};
\filldraw (1.6,-2) circle (1pt) node[below left]{${-1}$};
\filldraw (-1.4,-2) circle (1pt) node[below]{\small{$q_1^{-1}$}};
\filldraw (-3.2,-2) circle (1pt) node[below]{\small{$q^{-1}$}};
\filldraw (-0.5,-2) circle (1pt) node[below]{\small{$y_0^{-1}$}};
\filldraw (2.2,-2) circle (1pt) node[below]{\small{$y_0$}};

\end{tikzpicture}
\caption{Signature table of $\re g(z, \xi)$ for $\xi \in \mathcal I_\varepsilon$.} \label{fig:1}
\end{figure}
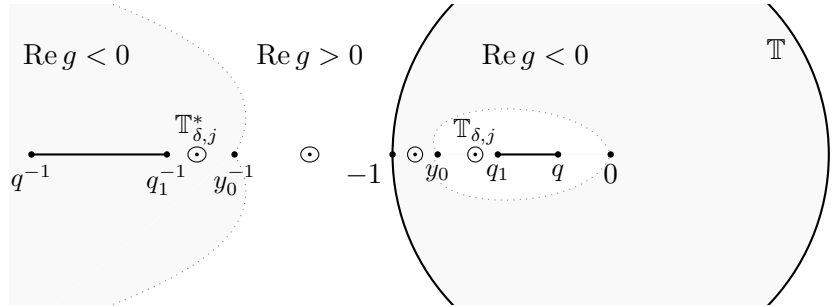

{\samepage\begin{Lemma}\label{propg} The function $g(z)$ satisfies the following properties
\begin{enumerate}[label=\rm{$(\alph*)$}]\itemsep=0pt
\item $g(z)$ is single valued on $\C \setminus \big[q^{-1}, q\big]$ and
$g\big(z^{-1}\big) = - g(z)$ for $z\in\C\setminus \big[q^{-1}, q\big]$,
\item $\re g(z) = 0$ for $z \in I \cup I^*\cup \{z\colon |z|=1\}$,
\item $g(q)=g\big(q^{-1}\big)=0$,
\item $g_-(z)=-g_+(z)$ for $z\in I \cup I^*$,
\item $\Phi(z)- g(z) = K(\xi) +O(z)$ as $z \to 0$, where\ $K(\xi)\in \R$ and
\begin{gather}\label{logcap}
\frac{\rm d}{{\rm d}\xi} K(\xi)=-\log\left(2\ti a\right).
\end{gather}
Here $\ti a$ is the logarithmic capacity of the set $\mathfrak S$, cf.~{\rm \cite[Lemma 5.4]{emt14}},
\item $g_+(z)-g_-(z)= - U-\xi \Lambda $ for $z \in J$, and

\noindent $g_+(z)-g_-(z)= U+\xi \Lambda $ for $J^*$,
where $U$ and $\Lambda$ are defined by~\eqref{defcycle}.
\end{enumerate}
\end{Lemma}
The signature table for $\re g(z,\xi)$ is given in Figure~\ref{fig:1}.

}

Recall that we enumerated the eigenvalues of the problem~\eqref{tl1},~\eqref{decay} starting from the gap $(b+2a, -1)$ in ascending order, i.e., $\la_1$ is the minimal eigenvalue and $\la_\aleph$ is the maximal one. The respective $z_j$ are enumerated in descending order.
Choose a small $\varepsilon_1>0$ such that
\begin{gather*}y_1(\xi_0 - \varepsilon_1)>z_1 + 2\delta, \qquad
y_2(\xi_{\aleph +1}+\varepsilon_1)<z_\aleph - 2\delta.
\end{gather*}
Smaller values of $\varepsilon$ or $\delta$ do not affect these conditions.
From Lemma \ref{discre} it follows that there are unique values $\xi_i\in (\xi_\aleph +\varepsilon_1, \xi_0 - \varepsilon_1)$ such that
\begin{gather*}
y_0(\xi_j)=z_j\qquad \forall j=1,\dots,\aleph.
\end{gather*}
Evidently, $\xi_{\aleph +1}<\dots<\xi_{j+1}<\xi_j<\dots<\xi_0$. We~denote
\begin{gather}\label{defIIps}
\mathcal I_\varepsilon:=[\xi_{\aleph+1} + \varepsilon, \xi_0 - \varepsilon]\setminus \bigcup\limits_{j=1}^\aleph (\xi_j-\varepsilon, \xi_j +\varepsilon)
=\bigcup_{j=1}^{\aleph +1} \mathcal I_\varepsilon^j,\qquad
\mathcal I_\varepsilon^j:=[\xi_{j}+\varepsilon, \xi_{j-1}-\varepsilon].
\end{gather}
From the considerations above it is straightforward to get the following
\begin{Lemma}\label{Ieps}
For any arbitrary small positive
$\varepsilon<\varepsilon_1$ one can choose $\delta>0$ such that for all
$\xi\in \mathcal I_\varepsilon$ the following inequalities are valid
\begin{gather}\label{defIeps}
\inf_{z_j\in\si_{\rm d}} \inf_{z\in \mathbb T_{\delta, j}}|\re g(z,\xi)| - |\Phi(z,\xi)-
\Phi(z_j,\xi)|> C(\varepsilon)>0.
\end{gather}
\end{Lemma}
Note that the infimum in~\eqref{defIeps} is taken along the circles around all points of the discrete spectrum $\si_{\rm d}$.
The VDO region
\begin{gather*}
\bigg \{(n,t)\in \Z\times\R_+\colon \frac{n}{t}\in \mathcal I_\varepsilon \bigg\}
\end{gather*}
consists of $\aleph+1$ nonintersecting sectors separated by arbitrary small sectors as depicted in~Fi\-gure~\ref{fig:2}.
If $\si_{\rm d}^{\rm gap}=\varnothing $, then $\aleph=0$ and the VDO region is the simply connected region
\begin{gather*}
\bigg \{(n,t)\in\Z\times \R_+\colon \frac{n}{t}\in [\xi_{\aleph+1} + \varepsilon,
\xi_0 - \varepsilon]\bigg\}.
\end{gather*}

For each $\xi\in\mathcal I_\varepsilon$ we divide the eigenvalues based on their relative location with respect
to the point $\mu_0(\xi)$ (respectively, $y_0(\xi)$) and introduce the Blaschke product
\begin{gather}\label{Pi1}
\Pi(z)= \Pi(z,\xi)= \prod_{y_0(\xi) < z_k<0} |z_k| \frac{z - z_k^{-1}}{z-z_k}.
\end{gather}
Note that
\begin{gather}\label{Pi}
\Pi\big(z^{-1}\big)=\Pi^{-1}(z),\qquad \Pi(0)>0.
\end{gather}
In fact,
\begin{gather}\label{zavis}
\Pi(z,\xi)=\Pi_j(z)= \prod_{z_j < z_k<0} |z_k| \frac{z - z_k^{-1}}{z-z_k} \qquad \text{for}\quad
\xi\in \mathcal I_\varepsilon^j.
\end{gather}
Set also (cf.~\eqref{circles})
\begin{gather}\label{geen}
E(z)= E(z,\xi) =\begin{cases}
\begin{pmatrix}1& \frac{z-z_j}{z_j
\gamma_j \E^{2t\Phi(z_j)} }\\ \frac{-z_j\gamma_j\E^{2t\Phi(z_j)}}{z-z_j} &1\end{pmatrix}, & z\in\mathbb D_{\delta, j}, \quad z_j \in (y_0, 0),\\
\si_1 E\big(z^{-1}\big)\si_1, & z\in\mathbb D_{\delta, j}^*, \quad z_j \in (y_0, 0),\\
\id, & z\in \C\setminus \bigcup_{z_j \in (y_0, 0)}\ol{\mathbb D_{\delta,j}\cup \mathbb D_{\delta,j}^* }.
\end{cases}
\end{gather}
The matrix $E$ is not an identity matrix only in small vicinities of those $z_j$ which lie in the domain where $ \re g(z,\xi)>0$.

From properties $(a)$, $(e)$ of Lemma \ref{propg}, property~\eqref{Pi} and~\eqref{geen} it follows that if a vector $m$ satisfies~\eqref{symcon} and~\eqref{norm}, so does the vector
\begin{gather}\label{josh}
m^{(1)}(z)= m(z)E(z)\big[\Pi(z)\E^{t(\Phi(z) - g(z))}\big]^{-\sigma_3},\qquad
z\in \C\setminus (\Sigma_\delta\cup\Sigma_\delta^*\cup J\cup J^*),
\end{gather}
 where $\sigma_3$ is the third Pauli matrix. A~straightforward computation using Theorem~\ref{RHini} and~Lem\-ma~\ref{propg} shows that if $m(z)$ satisfies~\eqref{minin},~\eqref{vini}, then $m^{(1)}(z)$ given by~\eqref{josh} solves the jump problem
 \begin{gather*}
 m_+^{(1)}(z)=m_-^{(1)}(z)v^{(1)}(z),\qquad
 z\in \Sigma_\delta\cup\Sigma_\delta^*\cup J \cup J^*,
 \end{gather*}
where
\begin{gather*}
v^{(1)}(z) = \begin{cases}
\begin{pmatrix}
\E^{t(g_+(z)-g_-(z))} & 0 \\
\Pi^{-2}(z)\chi(z) & \E^{-t(g_+(z)-g_-(z))}
\end{pmatrix}\!, & z \in I,
\\[3ex]
\begin{pmatrix}
\E^{t(g_+(z)-g_-(z))} & 0 \\
0 & \E^{-t(g_+(z)-g_-(z))}
\end{pmatrix}\!, & z \in J,
\\[3ex]
\begin{pmatrix}
1&0\\
\Pi^{-2}(z) R(z) \E^{2t g(z)} & 1
\end{pmatrix}\!, & z \in \mathcal C_\delta,
\\[3ex]
\mathcal A_j(z), & z\in\mathbb T_{\delta, j},\ z_j\notin (y_0, 0),
\\
\mathcal B_j(z), & z\in\mathbb T_{\delta, j},\ z_j\in (y_0, 0),
\\
\si_1 (v^{(1)}(z^{-1}))\si_1, & z\in \Sigma_\delta^*\cup J^*.
\end{cases}
\end{gather*}
Here
\begin{gather*}
 \mathcal A_j(z)= \mathcal A_j(z,\xi)=
 \begin{pmatrix} 1& 0\\ \frac{\gamma_j z_j}{\Pi^2(z)(z-z_j)}\E^{2tH_j(z)}&1\end{pmatrix}\!,
 \\
 \mathcal B_j(z)=\mathcal B_j(z,\xi)=
 \begin{pmatrix} 1& 0\\ \frac{\Pi^2(z)(z-z_j)}{\gamma_j z_j}\E^{-2tH_j(z)}&1\end{pmatrix}\!,
 \end{gather*}
and
\begin{gather*}
H_j(z)=H_j(z,\xi)= \Phi(z_j)-\Phi(z) +g(z,\xi).
\end{gather*}
With our choice of the VDO region we evidently have
\begin{Lemma}\label{ABj} Uniformly with respect to $\xi\in\mathcal I_\varepsilon$
\begin{gather} \label{HHj}
 \sup_{z_j\in\si_{\rm d}}\sup_{z\in \mathbb T_{\delta, j}}\left(\|\mathcal A_j(z)-\id\|
+ \|\mathcal B_j(z)-\id\|\right)\leq C_1(\varepsilon)\E^{-C(\varepsilon)t},
\\
\label{mathcalCeps}
 \sup_{z\in \mathcal C_\delta\cup \mathcal C_\delta^*}\big\| v^{(1)}(z)-\id\big\|\leq C_2(\varepsilon)\E^{-C(\varepsilon)t}.
\end{gather}
Here $\|\cdot\|$ is a norm of $2\times 2$ matrices.
\end{Lemma}
\begin{Remark}
The function $m^{(1)}(z)$ inherits the singularities of $m(z)$ described in Theorem \ref{RHini}.
\end{Remark}
Recall that these singularities essentially depend on the presence or absence of resonances at~points~\eqref{mathcalJ}.
In the next step we apply the lense mechanism around $I$ and $I^*$, which will at the same time weaken these singularities. We~will use $\ell=-1,0,1$ to indicate singularities as follows.
In $\C\setminus [q^{-1}, q]$ introduce a function $Q(z)$ such that
\begin{gather}\label{defQuu}
Q^4(z)=\begin{cases}
\frac{(z-q)(z- q_1)}{(zq_1 - 1)(zq -1)}, &\text{$q$ nonresonant},\quad \text{$q_1$ nonresonant}\ ( \ell=-1),
\\[1ex]
\frac{(zq_1 - 1)(zq -1)}{(z-q)(z- q_1)}, &\text{$q$ resonant}, \quad\phantom{non} \text{$q_1$ resonant}\ (\ell=1),
\\[1ex]
\frac{(zq_1 - 1)(z-q)}{(zq-1)(z- q_1)}, &\text{$q$ nonresonant},\quad \text{$q_1$ resonant}\ (\ell=0),
\\[1ex]
\frac{(zq - 1)(z-q_1)}{(zq_1-1)(z- q)}, &\text{$q$ resonant},\quad\phantom{non} \text{$q_1$ nonresonant}\ (\ell=0),
\end{cases}
\end{gather}
with the branch of the forth root defined by the condition $Q(1)=1$. Evidently,
$Q(z^{-1})=Q^{-1}(z)$ and $Q(0)>0$.
It has jumps on $I\cup I^*$ for $\ell=0$ and on $I\cup I^*\cup J\cup J^*$ for $\ell=\pm 1$.
The function
\begin{gather*}
\Omega(z,s)=\frac {1}{2 s}\frac{s+z}{s-z}
\end{gather*}
can be considered as the Cauchy kernel for symmetric contours, because $\Omega(z,s)=\frac{1}{z-s}(1 + o(1))$ as $z\to s$, and
\begin{gather*}
\Omega\big(z,s^{-1}\big)\,{\rm d}\big(s^{-1}\big)=\Omega\big(z^{-1},s\big)\,{\rm d}s.
\end{gather*}
Using this kernel allows us to preserve the symmetry condition. Set
\begin{gather}\label{defmatp}
\mathcal P(z) = \sqrt{(z-q)(z-q_1)\big(z-q_1^{-1}\big)\big(z-q^{-1}\big)z^{-2}}, \qquad
z \in \ol{\C\setminus (I\cup I^*)}.
\end{gather}
This function satisfies the symmetries
\begin{gather*}
\mathcal P(z^{-1})=\mathcal P(z)\quad \text{for}\quad z \in \C\setminus (I\cup I^*) \qquad \text{and}\qquad \mathcal P_-(z) = -\mathcal P_+(z)\quad \text{for}\quad z \in I \cup I^*.
\end{gather*}
	Define
\begin{gather*}
S(z) =\frac{1}{2\pi \I} \int_{q}^{q^{-1}} \Omega(z,s)f(s)\,{\rm d}s, \qquad
f(s) : = \begin{cases}
\frac{\log(\Pi^{-2}(s)Q^{-2}(z)|\chi(s)|)}{\mathcal P_+(s)}, & s \in I, \\
\frac{\I \ti\Delta}{\mathcal P(s)}, & s \in J,\\
f\big(s^{-1}\big), & s \in \big[ q^{-1}, -1\big],
\end{cases}
\end{gather*}
where
\begin{gather*}
\ti \Delta=\ti\Delta (\xi)= - \I \int_I \frac{\log\big(Q^{-2}(s)\Pi^{-2}(s)|\chi(s)|\big)}{\mathcal P_+(s)}\frac{{\rm d}s}{s}
\bigg(\int_{J} \frac{{\rm d}s}{s \mathcal P(s)} \bigg)^{-1}.
\end{gather*}
It is straightforward to verify that $S(z)$ solves the scalar RH problem
\begin{gather*}
\mathcal S_+(z) = \mathcal S_-(z) + f(z), \qquad z \in \big[q,q^{-1}\big],
\\
\mathcal S(z^{-1}) = - \mathcal S(z), \qquad z \in \C\setminus \big[q,q^{-1}\big],
\\
\mathcal S(z) = O(z), \qquad z\to 0.
\end{gather*}

The above considerations imply that the function
\begin{gather}\label{F(z)}
\mathcal F(z) = \E^{\mathcal P(z)\mathcal S(z)}, \qquad z \in \C \setminus \big[q^{-1},q\big],
\end{gather}
is the unique solution of the following RHP with jump along $I\cup I^*\cup J\cup J^*$,
\begin{enumerate}\itemsep=0pt
\item[$(i)$] $\mathcal F_+(z) \mathcal F_-(z) = \Pi^{-2}(z)|\chi(z)|Q^{-2}(z)$ for $z \in I$,
\item[$(ii)$] $\mathcal F_+(z) = \mathcal F_-(z) \E^{\I \ti\Delta}$ for $z \in J$,
\item[$(iii)$] $\mathcal F(z^{-1})=\mathcal F^{-1}(z)$ for $z \in \C \setminus \big[q^{-1}, q\big]$,
\item[$(iv)$] $\mathcal F(0)> 0$, $\mathcal F(1)=1$.
\end{enumerate}
Since $\chi(s)Q^{-2}(s)\neq 0 $ as $s\in I\cup I^*$ and it is a continuous function on $I\cup I^*$, then $\mathcal F(z)$ also has nonzero finite limiting values as $z\to \ti q\in\mathcal J$ (cf.~\cite{Muskh}). It~is straightforward to obtain
\begin{Lemma}\label{lemF}
The function $F(z):=\mathcal F(z)Q(z)$, defined for $z\in \C \setminus \big[q^{-1}, q\big]$ by~\eqref{defQuu}--\eqref{F(z)},
solves the following RHP
\begin{enumerate}\itemsep=0pt
\item[$(i)$] $F_+(z) F_-(z) = \Pi^{-2}(z)|\chi(z)|$ for $z \in I$,
\item[$(ii)$] $F_+(z) = F_-(z) \E^{\I \Delta}$ for $z \in J$,
\item[$(iii)$] $F\big(z^{-1}\big)=F^{-1}(z)$ for $z \in \C \setminus \big[q^{-1}, q\big]$,
\item[$(iv)$] $F(0)> 0$, $F(1)=1$,
\end{enumerate}
where
\begin{gather}\label{Delta}
\Delta=\Delta(\xi)=- \I \int_I \frac{\log\big(Q^{-2}(s)\Pi^{-2}(s,\xi)|\chi(s)|\big)}{\mathcal P_+(s)}\frac{{\rm d}s}{s}
\bigg(\int_{J} \frac{{\rm d}s}{s \mathcal P(s)} \bigg)^{-1} +\ell\pi.
\end{gather}
In a vicinity of $\ti q\in \{q, q_1\}$ we have $F(z)=C(z-\ti q)^{1/4}(1 + o(1))$ if $\ti q$ is a nonresonant point, and $F(z)=C(z-\ti q)^{-1/4}(1 + o(1))$ if $\ti q$ is a resonant point. The jumps of $F$ along the contours~$I^*$ and $J^*$ as well as its behavior at $q^{-1}, q_1^{-1}$ are uniquely defined by the symmetry $(iii)$.
\end{Lemma}
Note that the only dependence of $\Delta(\xi)$ on $\xi$ is due to the Blaschke product~\eqref{Pi1} depending on $\xi$. It~means that $\Delta(\xi)$ has constant values on every interval $\mathcal I_\varepsilon^j$. By~\eqref{zavis} we obtain that
\begin{gather}\label{dconst}
\Delta(\xi)=\Delta_j \qquad \text{for}\quad \xi\in \mathcal I_\varepsilon^j, \quad j=1,\dots,\aleph+1,
\end{gather}
where
\begin{gather}\label{Deltaj}
\Delta_j=- \I \int_I \frac{\log\big(Q^{-2}(s)\Pi^{-2}_j(s)|\chi(s)|\big)}{\mathcal P_+(s)}\frac{{\rm d}s}{s}
\left(\int_{J} \frac{{\rm d}s}{s \mathcal P(s)} \right)^{-1} +\ell\pi.
\end{gather}

Given these preparations, we can implement the next deformation step.

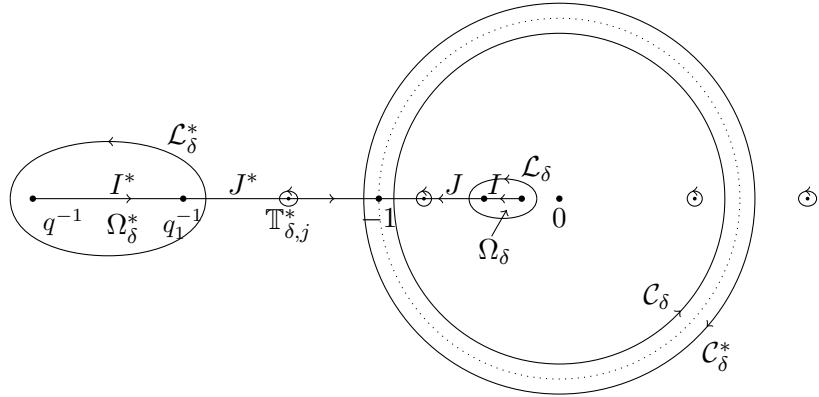
\begin{figure}[ht]
\centering
\begin{tikzpicture}

\draw[dotted] (0,0) circle (2.4cm);
\draw (0,0) circle (2.6cm);
\draw (0,0) circle (2.2cm);

\draw (-7,0) -- (-0.5,0);

\filldraw (0,0) circle (1pt) node[below]{$0$};
\filldraw (-0.5,0) circle (1pt);
\filldraw (-1,0) circle (1pt);
\filldraw (-2.4,0) circle (1pt) node[below]{$-1$};
\filldraw (-5,0) circle (1pt) node[below]{\small{$q_1^{-1}$}};
\filldraw (-7,0) circle (1pt) node[below right]{\small{$q^{-1}$}};

\draw[out=90, in=90] (-0.3,0) to (-1.2,0);
\draw[out=270, in=270] (-0.3,0) to (-1.2,0);
\draw[out=90, in=90] (-4.7,0) to (-7.3,0);
\draw[out=270, in=270] (-4.7,0) to (-7.3,0);

\draw [->] (-5.99,0.76) -- (-6,0.76);
\draw [->, thin] (-1.82,0.1) -- (-1.83,0.1);
\draw [->, thin] (-3.61,0.1) -- (-3.62,0.1);
\draw [->, thin] (3.28,0.1) -- (3.27,0.1);
\draw [->, thin] (1.78,0.1) -- (1.77,0.1);
\draw [->, thin] (1.6,-1.5) -- (1.61,-1.49);
\draw [->, thin] (1.97,-1.69) -- (1.96,-1.7);
\draw [->, thin] (-0.72,0.26) -- (-0.73,0.26);
\draw [->] (-5.7,0) -- (-5.69,0);
\draw [->] (-3,0) -- (-2.99,0);
\draw [->] (-0.78,0) -- (-0.79,0);
\draw [->] (-1.6,0) -- (-1.61,0);

\draw (3.3,0) ellipse (0.12cm and 0.1cm);
\filldraw (3.3,0) circle (0.5pt);
\draw (1.8,0) circle (0.1cm);
\filldraw (1.8,0) circle (0.5pt);
\draw (-1.8,0) circle (0.1cm);
\filldraw (-1.8,0) circle (0.5pt);
\draw (-3.6,0) ellipse (0.12cm and 0.1cm);
\filldraw (-3.6,0) circle (0.5pt) node[below]{$\T_{\delta,j}^*$};

\node at (1.3,-1.3) {$\mathcal C_\delta$};
\node at (2.1,-2.1) {$\mathcal C_\delta^*$};
\node at (-5.8,0.2) {$I^*$};
\node at (-5.8,-0.4) {$\Omega^*_\delta$};
\node at (-0.86,0.2) {$I$};
\node at (-1.4,0.2) {$J$};
\draw[->] (-0.9,-0.5) -- (-0.7,-0.15);
\node at (-0.86,-0.7) {$\Omega_\delta$} ;
\node at (-5,0.8) {$\mathcal L_\delta^*$};
\node at (-0.3,0.4) {$\mathcal L_\delta$};
\node at (-4.2,0.2) {$J^*$};

\end{tikzpicture}
\caption{Contour deformation of Step 2.} \label{fig:3}
\end{figure}

{\it Step 2}. Introduce two symmetric contours $\mathcal L_\delta$ and $\mathcal L_\delta^*=\big\{z\colon z^{-1}\in \mathcal L_\delta\big\}$ surrounding $I$ and $I^*$ counterclockwise at a small distance such that
\begin{gather}\label{condint}
\min_{z_j\in\si_{\rm d}} {\rm dist}\big(z_j, \ol\Omega_\delta\big)\gg \delta \qquad \text{and} \qquad
{\rm dist}\big(y_0(\xi),\ol\Omega_\delta\big)\gg \delta \qquad \forall \xi\in \mathcal I_\varepsilon,
\end{gather}
where $\Omega_\delta$ and $\Omega^*_\delta$ are the enclosed regions so that $\mathcal L_\delta=\partial \Omega_\delta$ and $\mathcal L^*_\delta=\partial \Omega^*_\delta$, see Figure~\ref{fig:3}.
Condition~\eqref{condint} ensures that $\mathcal L_\delta$ is away from the level line $\re g(.,\xi)=0$ and
from any point of the discrete spectrum.

Let $X(z)$, $z\in\Omega_\delta$, be the continuation of $\chi(z)$ as described in Remark \ref{continu}.
Set
\begin{gather*}
G^F(z) = \begin{pmatrix}
F^{-1}(z) & -\frac{\Pi^2 (z) F(z)}{X(z)} \E^{-2tg(z)}
\\
0 & F(z)
\end{pmatrix}\!,\qquad z\in \ol \Omega_\delta.
\end{gather*}
Define $m^{(2)}(z)$ by
\begin{gather*}
m^{(2)}(z)= \begin{cases}
m^{(1)}(z)G^F(z), & z \in \Omega_\delta,
\\
m^{(2)}\big(z^{-1}\big)\si_1, & z \in \Omega^*_\delta,
\\
m^{(1)}(z)(F(z))^{-\si_3}, & z \in \C \setminus (\Omega_\delta \cup \Omega^*_\delta).
\end{cases}
\end{gather*}


\begin{Theorem} \label{thm:v3}
For every $\xi \in \mathcal I_\varepsilon$, the vector function $m^{(2)}(z)=m^{(2)}(z,\xi)$ is the unique solution of the following RHP: to find a
holomorphic function in the domain \begin{gather*}\C\setminus(\Sigma_\delta\cup\Sigma_\delta^*\cup \mathcal L_\delta\cup\mathcal L_\delta^*\cup J\cup J^*)\end{gather*} which has continuous limits
on the sides of the contour $\Sigma_\delta\cup\Sigma_\delta^*\cup \mathcal L_\delta\cup\mathcal L_\delta^*\cup J\cup J^*$ except possibly at points $\mathcal J$~\eqref{mathcalJ} and satisfies:
\begin{itemize}
		\item the jump condition $m_{+}^{(2)}(z,n,t)=m_{-}^{(2)}(z,n,t) v^{(2)}(z,n,t)$,
\begin{gather}\label{v2}
v^{(2)}(z)=
\begin{cases}
v^{\rm mod}(z), &z\in I\cup I^*\cup J\cup J^*,
\\[.7ex]
\begin{pmatrix}
1 & \frac{\Pi^2(z)F^2(z)}{X(z)} \E^{-2tg(z)}
\\
0 & 1
\end{pmatrix}\!, &z \in \mathcal L_\delta,
\\[1.2ex]
[F(z)]^{-\sigma_3}v^{(1)}(z)[F(z)]^{\sigma_3}, &z\in \Sigma_\delta\setminus I,
\\		
\si_1 \big(v^{(2)}\big(z^{-1}\big)\big)\si_1, &z\in \Sigma_\delta^*\cup J^*\cup \mathcal L_\delta^*,
\end{cases}
\end{gather}
where $\Sigma_{\delta}$ is defined by~\eqref{sig} and
\begin{gather}\label{vmod} v^{\rm mod}(z)=
\begin{cases}
\I \si_1, &z \in I,
\\
\E^{-(n\Lambda+t U+ \I\Delta)\sigma_3}, &z \in J
\\
\sigma_1v^{\rm mod}\big(z^{-1}\big)\sigma_1, &z \in I^*\cup J^*,
\end{cases}
\end{gather}
\item the symmetry condition $m^{(2)}\big(z^{-1}\big) = m^{(2)}(z) \si_1$,
\item the normalization condition $m_1^{(2)}(0) \cdot m_2^{(2)}(0) = 1$, $m_1^{(2)}(0) > 0$,
\item at points of the set~\eqref{mathcalJ},
 $m^{(2)}(z)$ has at most a fourth root singularity,
 \begin{gather*}\label{sing4}
 m^{(2)}(z)=O(z-\kappa)^{-1/4},\qquad \text{as}\quad z\to \kappa\in \mathcal J.
\end{gather*}
\end{itemize}
\end{Theorem}

\begin{proof}
The proof of this theorem is completely analogous to the proof of \cite[Theorem 3.6]{empt19}
except for a small contour $I^\delta=[q_1 , q_1 -\delta]=\ol{\Omega_\delta}\cap [q_1, -1]\subset J$, where the jump matrix $v^{(2)}(z)=\big[G_-^F(z)\big]^{-1}v^{(1)}(z) G_+^F(z)$ should be evaluated.
 From Lemmas~\ref{propg}$(f)$, and~\ref{lemF}$(ii)$, it follows that
\begin{gather*}
v^{(2)}= \begin{pmatrix}
F_-& \frac{\Pi^2 F_-}{X} \E^{-2tg_-}
\\
0 & F_-^{-1}
\end{pmatrix}
\begin{pmatrix}
\E^{t(g_+ - g_-)} & 0\\ 0 & \E^{t(g_- - g_+)}
\end{pmatrix} \begin{pmatrix}
F_+^{-1} & -\frac{\Pi^2 F_+}{X} \E^{-2tg_+} \\
0 & F_+
\end{pmatrix}
\\ \hphantom{v^{(2)}}
{}= \begin{pmatrix} \frac{F_-}{F_+}\E^{t(g_+ - g_-)} & \frac{\Pi^2\E^{-t(g_+ + g_-)}}{X}(F_+F_--F_+F_-)
\\ 0 &\frac{F_+}{F_-}\E^{t(g_- - g_+)}\end{pmatrix}
=\begin{pmatrix}
\E^{-n\Lambda-t U- \I\Delta} & 0
\\
0 & \E^{n\Lambda + t U + \I\Delta}
\end{pmatrix}\!.
\end{gather*}
\looseness=1
On the symmetric contour $I^{\delta,*}$ (oriented from left to right) we get $v^{(2)}=v^{\rm mod}$ by the sym\-metry.
\end{proof}

\begin{Remark}\label{rem12}
According to~\eqref{Deltaj} and~\eqref{vmod} we see that
\begin{gather}\label{vej}
v^{\rm mod}(z)=v^{\rm mod}_j(z,n,t)
=\begin{cases}
\E^{-(n\Lambda + tU +\I\Delta_j)\sigma_3},& z\in J,\quad \frac{n}{t}\in \mathcal I^j_\varepsilon,
\\
\I\si_1, & z\in I, \\
\si_1 v^{\rm mod}\big(z^{-1}\big)\si_1,& z\in I^*\cup J^*.
\end{cases}
\end{gather}
\end{Remark}
Let us label the jump contour which appears in Theorem \ref{thm:v3} by
\begin{gather}\label{mathK}\mathcal K_\delta=\Sigma_\delta\cup\Sigma_\delta^*\cup \mathcal L_\delta\cup\mathcal L_\delta^*\cup J\cup J^*.
\end{gather}
We extend the matrix $v^{\rm mod}(z)$ to the whole contour $\mathcal K_\delta$ by defining it as
the identity matrix on the remaining part $\mathcal K_\delta\setminus (I\cup I^*\cup J\cup J^*)$.
\begin{Lemma}\label{concl} Uniformly with respect to $\xi\in \mathcal I_\varepsilon$,
\begin{gather}\label{important1}
\sup_{z\in\mathcal K_\delta}\big\|v^{(2)}(z) - v^{\rm mod}(z)\big\|\leq C_3(\varepsilon)\E^{-C_4(\varepsilon)t}, \qquad C_i(\varepsilon)>0.
\end{gather}
\end{Lemma}
\begin{proof}
Recall that
\begin{gather*}
\inf_{z\in\mathcal L_\delta}\re g(z,\xi)> C_5(\varepsilon)>0\qquad \forall \xi\in\mathcal I_\varepsilon.
\end{gather*}
This inequality verifies~\eqref{important1} on the contour $\mathcal L_\delta\cup\mathcal L_\delta^*$. Since the conjugation
\begin{gather*}[F(z)]^{-\sigma_3}v^{(1)}(z)[F(z)]^{\sigma_3}, \quad z\in \Sigma_\delta\setminus I,
\end{gather*}
does not impair estimates~\eqref{HHj},~\eqref{mathcalCeps} and since
$v^{(2)}(z)=v^{\rm mod}(z)$ for $z\in I\cup I^*\cup J\cup J^*$, then~\eqref{important1} is straightforward for the remaining part of $\mathcal K_\delta$.
\end{proof}

\section{Solution of the model problem and conclusive analysis}

In Section \ref{sec3}, Lemma \ref{lem:fg}, we constructed the vector-function $\ti m(p)$, which solves the jump problem~\eqref{jumpmod} for $\Delta$ given by~\eqref{jacinv}. One can treat this result in the following way: let $\Delta$ be an arbitrary real value and let $p_0$ be the unique solution of the Jacobi inversion problem~\eqref{jacinv}. Consider this point as the initial Dirichlet divisor and let $\hat a(n,t,\Delta)$, $\hat b(n,t,\Delta)$ be the finite gap solution associated with this divisor and with the spectrum $\mathfrak S$. In particular, we can construct $\ti m(p)$ associated with $\Delta$ given by~\eqref{Delta}. Being considered on the $z$-plane, the vector-function $\ti m(z)=\ti m(p(z))$ has additional jumps on $I\cup I^*$ due to~\eqref{tim} and solves the jump problem
\begin{gather*}\ti m_+(z)=\ti m_-(z)\begin{cases} \sigma_1, & z\in I\cup I^*,\\
v^{\rm mod}(z), & z\in J\cup J^*,\end{cases}
\end{gather*}
with $v^{\rm mod}(z)$ given by~\eqref{vmod} (or by~\eqref{vej}) on $J\cup J^*$.
Introduce the function
\begin{gather*}
H(z) = \sqrt[4]{\frac{(q_1-z)\big(q_1^{-1}-z\big)}{(q-z)\big(q^{-1}-z\big)}},
\end{gather*}
which satisfies
\begin{gather*}
H\big(z^{-1}\big)=H(z),\qquad z\in \C \setminus (I\cup I^*); \qquad H_+(z)=\I H_-(z), \qquad z\in I\cup I^*; \!\qquad H(0)=1.
\end{gather*}
Thus $m^{\rm mod}(z):=H(z)\ti m(z)$ is the unique solution of the following

\medskip

\noindent{\bf Model RH problem}.
{\it Find a vector-function $m^{\rm mod}(z)$ holomorphic in $\C\setminus \big[q, q^{-1}\big]$, continuous up to the boundary except of points of the set~\eqref{mathcalJ}, which satisfies the jump condition
\begin{gather*}
m_+^{\rm mod}(z)=m_-^{\rm mod}(z) v^{\rm mod}(z),\qquad z\in I\cup I^*\cup J\cup J^*,
\end{gather*}
with $v^{\rm mod}(z)$ given by~\eqref{vmod}, and the symmetry and normalization conditions
\begin{gather*}
m^{\rm mod}\big(z^{-1}\big)=m^{\rm mod}(z)\sigma_1, \qquad
 m^{\rm mod}_1(0)=\big[m^{\rm mod}_2(0)\big]^{-1}>0.
\end{gather*}
At points of the set~\eqref{mathcalJ} it has a fourth root singularity
\begin{gather*}
m^{\rm mod}(z)=O(z-\kappa)^{-1/4}, \qquad \text{as}\quad z\to\kappa\in\mathcal J.
\end{gather*}}
Uniqueness of the solution of such a problem was established in~\cite{emt14}.

The dependence of $m^{\rm mod}(z)$ on $n, t$ and $\Delta\big(\frac nt\big)$ is due to the jump
$\exp \big(\big(-n\Lambda -tU +\I\Delta\big(\frac nt\big)\big)\sigma_3\big)$. For~large $n$ and $t$, if $\frac{n}{t}\in \mathcal I_\varepsilon^j$, then $\frac{n+1}{t}\in \mathcal I_\varepsilon^j$. Recall that $\Delta(\xi)$ has constant values $\Delta_j$ on $\mathcal I_\varepsilon^j$ (cf.~\eqref{Deltaj}). Therefore,
\begin{gather*}
m^{\rm mod}(z,n,t,j):=m^{\rm mod}(z,n,t, \Delta_j)\quad\ \text{and} \quad\ m^{\rm mod}(z,n+1,t,j):=m^{\rm mod}(z,n+1,t, \Delta_j)
\end{gather*}
are well defined for $\frac{n}{t}\in \mathcal I_\varepsilon^j$. Here $m^{\rm mod}(z,n+1,t,j)$ is the solution of the jump problem with the jump $\exp\left(-((n+1)\Lambda + tU +\I\Delta_j)\sigma_3\right)$ on $J$ and associated jumps on $J^*\cup I\cup I^*$.
Since
\begin{gather*}
H^2(z)= \bigg(1-\frac{z}{2q_1}-\frac{zq_1}{2}+O\big(z^2\big)\bigg)\bigg(1+\frac{z}{2q}+\frac{zq}{2}+O\big(z^2\big)\bigg) = 1-2z(2a) + O\big(z^2\big)
\end{gather*}
and $\la=\frac{1}{2z} (1+o(1))$, then Theorem~\ref{theorRHs} and~\eqref{dconst} imply

\begin{Lemma}\label{lem:imp} For all $n\to\infty$, $t\to\infty$ and $\frac{n}{t}\in \mathcal I_\varepsilon^j$,
\begin{gather}
 \lim_{z\to 0} \frac{ m_1^{\rm mod}(z, n, t,j)}{ m_1^{\rm mod}(z, n+1, t,j)}=
\hat a(n,t,\Delta_j)\ti a^{-1},\nonumber
\\
\lim_{z\to 0}\frac{1}{2z}\big( m_1^{\rm mod}(z,n,t,j) m_2^{\rm mod}(z,n,t,j)-1\big)=\hat b(n,t,\Delta_j).\label{impfor}
\end{gather}
Here the phase $\Delta_j$ is defined by~\eqref{Deltaj},~\eqref{zavis} and $\ti a=\mathrm{Cap}\,\mathfrak S$.
\end{Lemma}
Our next task is to prove the following approximation

\begin{Theorem}\label{justif}
For all $n\to\infty$, $t\to\infty$ such that $\frac{n}{t}\in \mathcal I_\varepsilon^j$,
the following asymptotic holds as~$z\to 0$
\begin{gather}\label{serioz}
m^{(2)}(z,n,t)=m^{\rm mod}(z,n,t,j) + O\big(\E^{-C(\varepsilon)t}\big)(1 +O(z)), \qquad C(\varepsilon)>0.
\end{gather}
Moreover,
\begin{gather}
m_1(z)m_2(z)=m^{\rm mod}_1(z,n,t,j)m^{\rm mod}_2(z,n,t,j)
+ \beta_1^j(\xi, t) +\beta_2^j(\xi, t)z + \beta_3^j(\xi, t)O\big(z^2\big),\nonumber
\\ \hphantom{m_1(z)m_2(z)=}
z\to 0,\label{finalapr}
 \end{gather}
 where $\beta_k^j(\xi, t)=O\big(\E^{-C(\varepsilon)t}\big)$, $k=1,2,3$, uniformly with respect to $\xi\in\mathcal I_\varepsilon^j$.

\end{Theorem}
The proof of Theorem~\ref{justif} essentially repeats the arguments used in \cite[Section~7]{empt19}.
But since this theorem provides the justification of our asymptotic analysis, we will briefly describe the key points of the final analysis as applied to our case.

According to the standard approach (cf.~\cite{Bleher, its}), to perform the conclusive analysis we first have to evaluate in $\C$ the ``error vector"
\begin{gather}
\label{error}
m^{\rm err}(z,n,t,j):=m^{(2)}(z,n,t)\big[M^{\rm mod}_j(z,n,t)\big]^{-1},
\end{gather}
where $M^{\rm mod}_j(z,n,t)$ is a matrix solution of the model RHP. Recall that by Remark~\ref{rem12}, the model jump problem
\begin{gather}\label{Mmod}
M^{\rm mod}_{j,+}(z,n,t)= M^{\rm mod}_{j,-}(z,n,t)v^{\rm mod}_j(z,n,t), \qquad
z\in I\cup I^*\cup J\cup J^*, \qquad
\frac{n}{t}\in\mathcal I^j_\varepsilon,
\end{gather}
has a piecewise constant (with respect to $\xi$) jump matrix $v^{\rm mod}(z)$ given by~\eqref{vej}. The choice of such a matrix solution is not unique. However, a proper solution $M_j(z):=M_j^{\rm mod}(z,n,t)$ should be invertible and satisfy at least the symmetry $M_j\big(z^{-1}\big)=\sigma_1 M_j(z)\sigma_1$, because the vector $m^{\rm err}(z,n,t,j)$ should fulfill the standard symmetry property. Moreover, the singularities of $m^{\rm err}$ should be removable outside the jump contour and should not be more than $L^2$-integrable on the contour. By construction~\eqref{error}, $m^{\rm err}$ does not have a jump on $I\cup I^*\cup J\cup J^*$. Since $m^{(2)}(z)$ has singularities of order $O(z-\ti q)^{-1/4}$ at $\ti q \in\mathcal J$~\eqref{mathcalJ}, the matrix $[M_j(z)]^{-1}$ should have singularities of order not exceeding $o(z-\ti q)^{-3/4}$ at $\tilde q\in\mathcal J$ and should be less than poles at its other singular points outside the contour. As it is shown in~\cite{ept19} for the KdV shock wave (and the same is true for the Toda shock case) such an invertible solution with weak singularities does not exist for certain arbitrary large $t$ at any direction $\xi\in\mathcal I_\delta$. Instead, we choose a matrix solution with poles at points $1, -1$, such that its determinant is equal to 1 and $m^{\rm err}$ has removable singularities at~these points.

To explain the construction of our solution in more detail, recall that the quasimomentum~\eqref{struct} has the jump on $J\cup J^*$ defined by~\eqref{defcycle}. Introduce the function
\begin{gather*}
G(z):=\exp\bigg(\int_{b-2a}^{p(z)}\omega_{\infty_+,\infty_-}\bigg), \qquad z\in \C\setminus\big[q, q^{-1}\big].
\end{gather*}
Recall that for $z\in \mathcal D$ we have $p(z)=\big(\frac{z+z^{-1}}{2}, +\big)\in\M$ (upper sheet), and for $z\in \mathcal D^*$ we have $p(z)=\big(\frac{z+z^{-1}}{2}, -\big)$. Then $G(z)$ has the following properties~\cite[Section~5]{empt19}:
\begin{itemize}
\item $G(z)$ is holomorphic on $\C\setminus\big[q, q^{-1}\big]$ and satisfies $G\big(z^{-1}\big)= G^{-1}(z)$.
\item Its jumps are given by
\begin{gather*}
G_+(z) = G_-(z)\E^{-\Lambda},\qquad z\in J,
\\
G_+(z) = G_-(z)\E^{\Lambda}, \qquad z\in J^*,
\\
G_\pm(z) = \big[G_\pm\big(z^{-1}\big)\big]^{-1}, \qquad z\in I\cup I^*.
\end{gather*}
\item The following asymptotic expansion is valid,
\begin{gather*}
G(z)=-\frac{\ti a}{2 z}\big(1 + 2\ti b z+ O\big(z^2\big)\big), \qquad
G(z^{-1})=-\frac{2z}{\ti a}\big(1 -2\ti b z+O\big(z^2\big)\big),
\end{gather*}
where $\ti a=\mathrm{Cap}\,\mathfrak S$ and $\ti b\in\R$.
\end{itemize}

\begin{Lemma}
Introduce the following vector function holomorphic in $\C\setminus ([q, q^{-1}]\cup\{0\})$,
\begin{gather*}
m^\#(z,n,t,j)=m^{\rm mod}(z,n+1,t,j)[G(z)]^{-\sigma_3}.
\end{gather*}
Then $m^\#(z)$ solves the jump problem
\begin{gather*}
m^\#_+(z,n,t,j)=m^\#_-(z,n,t,j)v^{\rm mod}_j(z,n,t), \qquad z\in I\cup I^*\cup J\cup J^*, \qquad \frac{n}{t}\in\mathcal I^j_\varepsilon,
\end{gather*}
where $v^{\rm mod}_j(z,n,t)$ is given by~\eqref{vej}.
The vector $m^\#(z)$ satisfies the symmetry condition $m^\#_2\big(z^{-1}\big) = m^\#_1(z)$.
The normalization condition is not fulfilled, instead we have
\begin{gather*}
 m^\#_1(z,n,t,j)=-\frac{2z}{\ti a}m^{\rm mod}_1(0,n+1,t,j)(1+ O(z)),
 \\
 m^\#_2(z,n,t,j)=-\frac{\ti a}{2 z}m^{\rm mod}_1(\infty,n+1,t,j)(1 + O(z)),\qquad
 \text{as}\quad z\to 0.
\end{gather*}
\end{Lemma}
\begin{proof} The proof follows immediately from the properties of $G$ and $m^{\rm mod}$ above.
\end{proof}

The function
\begin{gather*}\rho(z)=\rho(z,n,t,j)=\frac{m^{\rm mod}_2(0,n,t,j)m^{\rm mod}_2(\infty,n+1,t,j)}{2\ti a\big(z^{-1} - z\big)}
\end{gather*} is defined for all $z\neq\pm 1$ and is odd,
$\rho(z)=-\rho\big(z^{-1}\big)$.
The function $\rho$ does not have jumps, therefore the vector $\rho(z)m^\#(z)$ solves the same jump problem with~\eqref{vej}. However, it is bounded as $z\to 0$, $z\to \infty$ and has simple poles at~$1$ and~$-1$ instead.
In conclusion, the vector
\begin{gather*}
\Psi(z,n,t,j)=\frac{1}{2}m^{\rm mod}(z,n,t,j)+\rho(z)m^\#(z,n,t,j)
\end{gather*}
solves our vector model RHP, and the same is true for $\Psi\big(z^{-1},n,t,j\big)\sigma_1$.
From here on, we fix the parameters $n$, $t$, $j$ and omit them to shorten notations when necessary. In particular, the symmetry conditions for $m^{\rm mod}$ and $m^\#$ and oddness of $\rho$ imply for the components of
 $\Psi$ that
\begin{gather}
\Psi_1(z)=\frac{1}{2}m_1^{\rm mod}(z) +\rho(z)m^\#_1(z), \qquad \Psi_2(z)=\frac{1}{2}m_2^{\rm mod}(z)+\rho(z)m^\#_2(z),\nonumber
\\
\Psi_1\big(z^{-1}\big)=\frac{1}{2}m_2^{\rm mod}(z)-\rho(z)m^\#_2(z),\quad \Psi_2\big(z^{-1}\big)=\frac{1}{2}m_1^{\rm mod}(z)-\rho(z)m^\#_1(z).\label{defPsi}
\end{gather}

\begin{Lemma}[{\cite[Lemma 5.2]{empt19}}] \label{aboutmatrix}\quad
\begin{enumerate}\itemsep=0pt
\item[$(i)$] The matrix $M^{\rm mod}(z)=M^{\rm mod}_j(z,n,t)$
\begin{gather*}
M^{\rm mod}(z)=\begin{pmatrix} \Psi_1(z) & \Psi_2(z)\\ \Psi_2(z^{-1}) & \Psi_1(z^{-1})\end{pmatrix}\!, \qquad z\in \C\setminus \big[q^{-1},q \big],
\end{gather*}
is a meromorphic matrix solution for the model jump problem
\begin{gather*}
M^{\rm mod}_+(z)= M^{\rm mod}_-(z)v^{\rm mod}(z),\qquad z\in I\cup I^*\cup J\cup J^*,
\end{gather*}
with $v^{\rm mod}(z)$ given by~\eqref{vmod}. It~has simple poles at $z=\pm 1$.
\item[$(ii)$] $M^{\rm mod}(z)$ satisfies the symmetry
\begin{gather}\label{symmat}
M^{\rm mod}(z^{-1})=\sigma_1 M^{\rm mod}(z)\sigma_1.
\end{gather}
\item[$(iii)$] The following equality is valid,
\begin{gather}\label{zavis2}
m^{\rm mod}(z,n,t)=(1, 1)M^{\rm mod}(z,n,t).
\end{gather}
\item[$(iv)$] The determinant of $M^{\rm mod}(z)$ is a constant with respect to $n$, $t$, $j$, $z$,
\begin{gather}\label{determ}
\det M^{\rm mod}(z)=1,\qquad z\in \C.
\end{gather}\end{enumerate}
\end{Lemma}
Using this lemma, we can establish the properties of the error vector function~\eqref{error}, which we denote here by $m^{\rm err}(z)$. We~recall the definition of $\mathcal K_\delta$ from~\eqref{mathK}.
\begin{Theorem}
The vector function $m^{\rm err}(z)$ is holomorphic in $\C\setminus \Xi_\delta$, where
\begin{gather*} \Xi_\delta:=\mathcal L_\delta\cup\mathcal L_\delta^*\cup C_\delta\cup C_\delta^*\cup\bigcup_{z_k\in\si_{\rm d}}\left(\mathbb T_{\delta,k} \cup \mathbb T_{\delta,k}^*\right)=\mathcal K_\delta\setminus (I\cup I^*\cup J\cup J^*),
\end{gather*}
and satisfies the following properties:
\begin{enumerate}\itemsep=0pt
\item[$(i)$] $m^{\rm err}(z)$ has removable singularities at $0$, $1$, $-1$, $\infty $ and on $\mathcal J$.
\item[$(ii)$] It solves the jump problem
\begin{gather}\label{merr}
m^{\rm err}_+(z)=m^{\rm err}_-(z)(\id + W(z)),\qquad z\in \Xi_\delta,
\end{gather}
 where
 \begin{gather}\label{errw}
 W(z)=M^{\rm mod}(z)\big(v^{(2)}(z) - \id\big)\big[M^{\rm mod}(z)\big]^{-1}.
 \end{gather}
\item[$(iii)$] It satisfies the symmetry condition
\begin{gather}\label{symer}
m^{\rm err}(z)=m^{\rm err}\big(z^{-1}\big)\sigma_1, \qquad z\in\C\setminus\Xi_\delta.
\end{gather}
In particular,
\begin{gather}\label{symcontur}
m_-^{\rm err}\big(z^{-1}\big)=m_-^{\rm err}(z)\si_1,\qquad z\in \Xi_\delta,
\\
\label{00}
m^{\rm err}_2(0)=m_1^{\rm err}(\infty)=\frac 12\bigg(\frac{m^{(2)}_1(0)}{m^{\rm mod}_1(0)} +\frac{m^{\rm mod}_1(0))}{m^{(2)}_1(0)} \bigg):=\tau>0.
\end{gather}
Here $m^{\rm mod}_1(0)=m^{\rm mod}_1(0,n,t,j)$ and $m^{(2)}(0)= m^{(2)}(0,n,t)$.
\end{enumerate}
\end{Theorem}
\begin{proof} The absence of singularities at the points $0$, $1$, $-1$, $\infty $ and on the set $\mathcal J$ was proven in~\cite[Lemmas~5.4 and~5.5]{empt19}. The jump~\eqref{merr} and the symmetry property~\eqref{symer} follow from~\eqref{v2}, \eqref{vmod}, \eqref{Mmod}, \eqref{symmat} and~\eqref{determ}. Property~\eqref{symcontur} holds due to the mutual orientation of the symmetric parts of the contour $\Xi_\delta$. Last,~\eqref{00} follows from~\eqref{error},~\eqref{defPsi},~\eqref{determ} and the definition of~$\rho$ which implies
\begin{gather*}
\lim_{z\to 0}\rho(z)m_2^\#(z)=-\frac{1}{m_1^{\rm mod}(0)}, \qquad \lim_{z\to 0}\rho(z)m_1^\#(z)=0.
\tag*{\qed}
\end{gather*}
\renewcommand{\qed}{}
\end{proof}

Now we are ready to prove Theorem \ref{justif}. We~follow the well-known approach via singular integral equations (see, e.g.,~\cite{dz}, \cite[Chapter~4]{its},~\cite{KTb}). A~peculiarity of this approach applied to the Toda equation is generated by the type of normalization condition of the vector RHP and the symmetry condition.
In particular, if we want to preserve the symmetry condition~\eqref{symer} in the Cauchy-type formula for
$m^{\rm err}(z)$, we should use a matrix Cauchy kernel (cf. \cite[equation~(B.8)]{KTb},
\begin{gather*}
\hat \Omega(s,z)
=\begin{pmatrix}\frac{1}{s-z} & 0\\ 0 & \frac{1}{s-z} -\frac{1}{s}\end{pmatrix}{\rm d}s,
\qquad s\in \Xi_\delta, \qquad z\notin \Xi_\delta.
\end{gather*}
Since~\cite[equation~(B.9)]{KTb}
\begin{gather*}
\hat\Omega\big(s, z^{-1}\big)=\si_1\hat \Omega\big(s^{-1},z\big)\si_1,
\end{gather*}
and $ W(s^{-1})=\si_1 W(s)\si_1,$ $ s\in \Xi_\delta,$
this implies with~\eqref{symcontur} and the orientation of $\Xi_\delta$ that the symmetry property holds:
\begin{gather*}\int_{\Xi_\delta}m^{\rm err}_-(s)W(s)\hat\Omega(s,z)= \int_{\Xi_\delta}m^{\rm err}_-(s)W(s)\hat \Omega\big(s,z^{-1}\big)\si_1.\end{gather*}
Note that the $1,1$-entry of the Cauchy kernel $\hat\Omega(s,z)$ has a zero at $z=\infty$ while the $2,2$-entry has a zero at $z=0$. From~\eqref{merr} and~\eqref{00} it follows that
\begin{gather*}
m^{\rm err}(z)=\big(m_1^{\rm err}(\infty), m_2^{\rm err}(0)\big) + \frac{1}{2\pi\I}\int_{\Xi_\delta}m^{\rm err}_-(s)W(s)\hat\Omega(s,z)
\\ \hphantom{m^{\rm err}(z)}
{}=\tau (1, 1) + \frac{1}{2\pi\I}\int_{\Xi_\delta}m^{\rm err}_-(s)W(s)\hat\Omega(s,z).
\end{gather*}
 Recall that according to Lemma~\ref{concl},
\begin{gather*}
\|W(z)\|=O\big(\E^{-C(\delta)t}\big),\qquad \text{uniformly with respect to}\quad
z\in \Xi_\delta\quad \text{and}\quad n,t,j\colon \frac{n}{t}\in \mathcal I^j_\varepsilon.
\end{gather*}
In turn, it implies the following
 \begin{Lemma}\label{lemimp}
Uniformly with respect to $\frac{n}{t}\in\mathcal I_\varepsilon$,
\begin{gather}
\label{2w}
\big\|z^k W(z) \big\|_{L^p(\Xi_\delta)}= O\big(\E^{-C(\delta)t}\big), \qquad
p \in [1,\infty], \qquad k = 0,1.
\end{gather}
\end{Lemma}

Now we are ready to apply the technique of singular integral equations.
Let $\mathfrak C$ denote the Cauchy operator associated with $\Xi_\delta$,
\begin{gather*}
(\mathfrak C h)(z)=\frac{1}{2\pi\I}\int_{\Xi_\delta}h(s)\hat\Omega(s,z), \qquad s\in\C\setminus\Xi_\delta,
\end{gather*}
where $h= \begin{pmatrix} h_1 & h_2 \end{pmatrix}\in L^2(\Xi_\delta)$ and satisfies the symmetry $h(s)=h\big(s^{-1}\big)\si_1$.
Let $(\mathfrak C_+ h)(z)$ and $(\mathfrak C_- h)(z)$ be the non-tangential limiting values of $ (\mathfrak C h)(z)$ from
 the left and right sides of $\Xi_\delta$, respectively.
As usual, we introduce the operator $\mathfrak C_{W}\colon L^2(\Xi_\delta)\cap L^\infty(\Xi_\delta)\to
L^2(\Xi_\delta)$ by $\mathfrak C_{W} h=\mathfrak C_-(h W)$, where $W$ is the error matrix~\eqref{errw}.
Then
\begin{gather*}
\|\mathfrak C_{W}\|=\|\mathfrak C_{W}\|_{L^2(\Xi_\delta)\to L^2(\Xi_\delta)}\leq C\|W\|_{L^\infty(\Xi_\delta)}=
O\big(\E^{-C(\varepsilon)t}\big),
\end{gather*}
as well as
\begin{gather}\label{6w}
\big\|(\id - \mathfrak C_{W})^{-1}\big\|=\big\|(\id - \mathfrak C_{W})^{-1}\big\|_{L^2(\Xi_\delta)\to L^2(\Xi_\delta)}\leq \frac{1}{1-O\big(\E^{-C(\varepsilon)t}\big)}
\end{gather}
for sufficiently large $t$. Consequently, for $t\gg 1$, on $\Xi_\delta$ we define a vector function
\begin{gather*}
\mu(s) =(\tau,\tau) + (\id - \mathfrak C_{W})^{-1}\mathfrak C_{W}\big((\tau,\tau)\big)(s),
\end{gather*}
with $\tau$ given by~\eqref{00}.
Then by~\eqref{2w} and~\eqref{6w}
\begin{gather}
\|\mu(s) - (\tau,\tau)\|_{L^2(\Xi_\delta)} \leq \big\|(\id - \mathfrak C_{W})^{-1}\big\| \|\mathfrak C_{-}\| \|W\|_{L^\infty(\Xi_\delta)}
= O\big(\E^{-C(\varepsilon)t}\big).\label{estmu}
\end{gather}
With the help of $\mu$, the vector function $m^{\rm err}(z)$ can be represented as
\begin{gather*}
m^{\rm err}(z)=(\tau,\tau) +\frac{1}{2\pi\I}\int_{\Xi_\delta}\mu(s) W(s) \hat \Omega(s,z),
\end{gather*}
and by virtue of~\eqref{estmu} and Lemma \ref{lemimp} we obtain as $z\to 0$
\begin{gather}\label{nush}
m^{\rm err}(z)=(\tau,\tau) + \frac{1}{2\pi\I } \int_{\Xi_\delta} (\tau,\tau) W(s)\hat\Omega(s,z) + E(z).
\end{gather}
Here $E(z)$ is a vector function holomorphic in a vicinity of $z=0$ which admits the estimate
\begin{gather*}
\|E(z)\|\leq \|W\|_{L^2(\Xi_\delta)}\|\mu (s)- (\tau,\tau)\|_{L^2(\Xi_\delta)}(1 +O(z))=O\big(\E^{-C(\varepsilon)t}\big)(1 +O(z)),
\end{gather*}
and $O(z)$ is uniformly bounded for $\frac{n}{t}\in\mathcal I_\varepsilon^j$.
From~\eqref{nush} and~\eqref{zavis2} we get
\begin{gather*}
m^{(2)}(z) = m^{\rm err}(z) M^{\rm mod}(z)=
\tau m^{\rm mod}(z) + \tau O\big(\E^{-C(\varepsilon)t}\big) E_1(z),
\end{gather*}
where $E_1(z)$ is a holomorphic vector function in a vicinity of $z=0$, uniformly bounded with respect to $n$, $t$, $j$ as $\frac{n}{t}\in \mathcal I_\varepsilon^j$.
The normalization conditions for $m^{(2)}$ and $m^{\rm mod}$ imply that
\begin{gather*}
\tau^2\big(1 + O\big(\E^{-C(\varepsilon)t}\big)\big)=1, \qquad \text{that is}, \qquad
\tau=1 + O\big(\E^{-C(\varepsilon)t}\big).
\end{gather*}
With~\eqref{00} this yields~\eqref{serioz} and at the same time~\eqref{finalapr}, which
 proves Theorem~\ref{justif}.

To finish the proof of Theorem \ref{theor:main} recall that in a vicinity of $z=0$ the initial vector-function~\eqref{m_ini} and the transformed function $m^{(2)}(z)$ are connected by
\begin{gather}\label{nadoe}
m^{(2)}(z,n,t)=m(z,n,t)\big[\Pi(z,\xi)\E^{t(\Phi(z,\xi) - g(z,\xi))}F(z,\xi)\big]^{-\sigma_3},\qquad \xi=\frac{n}{t}.
\end{gather}
 The results of Theorem \ref{justif} and Lemmas \ref{lem:as} and \ref{lem:imp} imply
\begin{gather*}
b(n,t)=\hat b(n,t,\Delta_j)+O\big(\E^{-C(\varepsilon)t}\big), \qquad
\frac{n}{t}\in \mathcal I_\varepsilon^j.
\end{gather*}
On the other hand, from~\eqref{asympm},~\eqref{nadoe} and Lemma \ref{propg}(e), it follows
\begin{gather*}
2a(n,t)=\frac{m_1(0,n,t)}{m_1(0,n+1,t)}= \frac{m_1^{(2)}(0,n,t)}{m_1^{(2)}(0,n+1,t)}
\E^{t\left(K\left(\frac nt\right) - K\left(\frac{n+1}{ t}\right)\right)},
\end{gather*}
because $\Pi(z)$ and $F(z)$ are the same for $n$ and $n+1$.
By~\eqref{logcap}, $K(\xi)=-\log(2\ti a)(\xi - {\rm const})$, that is,
\begin{gather*}
t\bigg(\!K\bigg(\frac{n}{t}\bigg) - K\bigg(\frac{n+1}{ t}\bigg)\!\bigg)=\log(2\ti a).
\end{gather*}
Using~\eqref{impfor} we finally get
\begin{gather*}
2a(n,t)= \frac{m_1^{\rm mod}(0,n,t)}{m_1^{\rm mod}(0,n+1,t)}\E^{2\ti a}+ O\big(\E^{-C(\varepsilon)t}\big)=2\hat a(n,t,\Delta_j)+O\big(\E^{-C(\varepsilon)t}\big).
\end{gather*}
This finishes the proof of our main result, Theorem \ref{theor:main}.

\subsection*{Acknowledgements}
This research was supported by the Austrian Science Fund (FWF) under Grant No.~P31651. We~thank the referees for their careful reading and their recommendations.


\pdfbookmark[1]{References}{ref}
\LastPageEnding

\end{document}